\documentclass[11pt, a4paper]{article}
    \usepackage{mathtools,amsthm,amssymb,amsfonts,dsfont}
    \usepackage{amsmath}
    \usepackage{pgfplots}
    \usepackage[authoryear,round,longnamesfirst]{natbib}
\bibpunct{(}{)}{;}{a}{,}{,} 
    \usepackage{eufrak}
    \usepackage[normalem]{ulem}
    \usepackage{appendix}
    \usepackage{pgfplots}
        \pgfplotsset{width=12cm,compat=1.9}
        \usepackage{pst-func}
        \psset{unit=2cm}
        \usepgfplotslibrary{fillbetween}
    \usepackage{hyperref}
    \usetikzlibrary{shapes, patterns, arrows.meta, positioning}
\hypersetup{colorlinks=true,urlcolor=blue,citecolor=blue,linkcolor=red}
    \usepackage[center]{caption}
    \usepackage{subcaption}
    \usepackage{ragged2e}
	\usepackage{comment}
    \usepackage{float}
    \usepackage{array}  
    \newcolumntype{C}[1]{>{\centering\arraybackslash}p{#1}}
	\usepackage{setspace}
	    \onehalfspace
	\usepackage{bigints}
	\usepackage{booktabs}
	\usepackage{epsfig}
	\usepackage{graphics}
    \usepackage[flushleft]{threeparttable}
    \usepackage{multirow}
    
    \usepackage{adjustbox}
	\usepackage{lscape}
    \usepackage{lipsum}
 \usepackage{longtable}
	\usepackage[english]{babel}
	\usepackage{color}
	\usepackage{marvosym}
	\usepackage[margin=1in]{geometry}
	\usepackage{tikz}
    \usepackage{tgpagella}
	\usepackage{chngcntr}
    \usepackage{apptools}
    \usepackage{tikz, smartdiagram}
     \usetikzlibrary{shapes.multipart}  
\usetikzlibrary{decorations.pathreplacing,calligraphy}
     \tikzset{
         state/.style={rectangle split, draw=black, text width=3cm}
     }
     \tikzset{every picture/.append style={remember picture}}

    \AtAppendix{\counterwithin{assumption}{section}}
    \AtAppendix{\counterwithin{table}{section}}
    \AtAppendix{\counterwithin{lem}{section}}
    \AtAppendix{\counterwithin{corollary}{section}}
    \AtAppendix{\counterwithin{rem}{section}}
    \AtAppendix{\counterwithin{example}{section}}
    \AtAppendix{\counterwithin{figure}{section}}
    \AtAppendix{\counterwithin{thm}{section}}
    \AtAppendix{\counterwithin{prop}{section}}
	
	\DeclareMathOperator*{\argmax}{arg\,max}
	\DeclareMathOperator*{\argmin}{arg\,min}

    \newcommand{\indic}{\mathbb I}
    
    \newcommand{\bp}{\mathbf{p}} 
    \newcommand{\bx}{\mathbf{x}} 
    \newcommand{\bq}{\mathbf{q}} 
    \newcommand{\bh}{\mathbf{h}} 
    \newcommand{\bt}{\mathbf{t}} 
    \newcommand{\bM}{\mathbf{M}}

    \newcommand{\wtq}{\overline{q}} 
    \newcommand{\wtomega}{\overline{\omega}} 
    
    \newcommand{\Cov}{\text{Cov}} 
     
    \newcommand{\myvec}{\text{vec}}

    \newtheoremstyle{indented}{20pt}{20pt}{\addtolength{\leftskip}{0.0em}}{}{\bfseries}{.}{.5em}{}
    \newtheoremstyle{indented_nonbold}{15pt}{15pt}{\addtolength{\leftskip}{1.5em}}{}{\itshape}{.}{.5em}{}

    \theoremstyle{indented}
    \theoremstyle{indented}\newtheorem {prop}{Proposition}
    \theoremstyle{plain}\newtheorem {thm}{Theorem}
    \theoremstyle{plain}
    \newtheorem*{thm*}{Theorem}
    \theoremstyle{indented}
    \theoremstyle{indented}\newtheorem {lem}{Lemma}
    \theoremstyle{indented}
    \theoremstyle{indented}
    \theoremstyle{indented}
    \theoremstyle{indented} 
    \theoremstyle{remark} 
    \theoremstyle{indented} \newtheorem{rem}{Remark}
    \theoremstyle{remark} 

    \title{Consumer Welfare \\ Under Individual Heterogeneity}
    
    \author{
    Charles Gauthier\thanks{KU Leuven; \href{mailto:charles.gauthier@kuleuven.be} {\tt charles.gauthier@kuleuven.be}}\quad
    Sebastiaan Maes\thanks{University of Antwerp; \href{mailto:sebastiaan.maes@uantwerpen.be} {\tt sebastiaan.maes@uantwerpen.be}}\quad Raghav Malhotra\thanks{University of Leicester; \href{mailto:r.malhotra@leicester.ac.uk} {\tt r.malhotra@leicester.ac.uk} 
    \newline We are grateful to Robert Akerlof, Roy Allen, David Baqaee, Dan Bernhardt, Debopam Bhattacharya, Pablo Becker, Luis Candelaria, Costas Cavounidis, Laurens Cherchye, Daniele Condorelli, Sam Cosaert, Ian Crawford, Andr\'e Decoster, Liebrecht De Sadeleer, Geert Dhaene, Peter Hammond, Stefan Hoderlein, Yuichi Kitamura, Danial Lashkari, Arthur Lewbel, Kenichi Nagasawa, Krishna Pendakur, Herakles Polemarchakis, Eric Renault, Camilla Roncoroni, Vladimir Smirnyagin, and Ao Wang. We also thank workshop and conference participants at Antwerp, Brussels, Leuven, Oxford, Tilburg, Warwick, AEE, ESEM, ECORES, ESCOE, EWMES, and LAGV. Sebastiaan benefited from fellowships from the Research Foundation Flanders (grants 11F8919N and 12C8623N). The results and their interpretation are the authors' sole responsibility. Researcher(s)' own analyses calculated (or derived) based in part on data from Nielsen Consumer LLC and marketing databases provided through the NielsenIQ Datasets at the Kilts Center for Marketing Data Center at The University of Chicago Booth School of Business. The conclusions drawn from the NielsenIQ data are those of the researchers and do not reflect the views of NielsenIQ. NielsenIQ is not responsible for, had no role in, and was not involved in analyzing and preparing the results reported herein.}}
    
    \date{\today\bigskip}
  
\begin{document}	
    \begin{titlepage}
        \maketitle
        \thispagestyle{empty}
          \begin{abstract}
            We propose a nonparametric method for estimating the distribution of consumer welfare from cross-sectional data with no restrictions on individual preferences. First demonstrating that moments of demand identify the curvature of the expenditure function, we use these moments to approximate money-metric welfare measures. Our approach captures both nonhomotheticity and heterogeneity in preferences in the behavioral responses to price changes. We apply our method to US household scanner data to evaluate the impacts of the price shock between December $2020$ and $2021$ on the cost-of-living index. We document substantial heterogeneity in welfare losses within and across demographic groups. For most groups, a naive measure of consumer welfare would significantly underestimate the welfare loss. By decomposing the behavioral responses into the components arising from nonhomotheticity and heterogeneity in preferences, we find that both factors are essential for accurate welfare measurement, with heterogeneity contributing more substantially.
        \end{abstract}
        
        
    \end{titlepage}

\section{Introduction}

Estimating consumer welfare under price changes is essential for policy evaluation across a wide range of domains, including tax reforms, trade liberalization, market interventions, and competition policies. The recent pandemic and ensuing large and sustained inflationary episodes have further renewed interest in measuring consumer welfare, highlighting the importance of accurate cost-of-living indices (e.g., \citeauthor*{fajgelbaum2016measuring}, \citeyear{fajgelbaum2016measuring}; \citeauthor{baqaee2022new}, \citeyear{baqaee2022new}; \citeauthor{atkinetal23}, \citeyear{atkinetal23}; \citeauthor{jaravel2023measuring}, \citeyear{jaravel2023measuring}). 

While these advances emphasize the nonhomotheticity of preferences, they often abstract from unobserved preference heterogeneity across households. Indeed, most datasets are cross-sectional and thus offer very limited behavioral insight under individual heterogeneity. In this paper, we introduce a novel nonparametric approach to estimating consumer welfare that is applicable in cross sections of consumers with data on prices and expenditures while accounting for unrestricted unobserved preference heterogeneity.

Our method uses moments of demand to identify local approximations of moments of welfare changes, thereby providing a flexible alternative to traditional parametric models.\footnote{Standard approaches impose parametric restrictions on preferences \citep{deatonmuellbauer, lewbelpendakur}. These work well when preferences are homogeneous but can misrepresent welfare impacts when unobserved heterogeneity is unrestricted. Accounting for preference heterogeneity is crucial in empirical applications, as traditional microeconometric models typically explain only a small fraction of the variation in consumer demand.} We apply our framework to measure changes in cost-of-living indices (CLIs) in the United States following the recent pandemic. Since our analysis accurately captures the distributional welfare impacts of price changes, it can offer valuable insights for policies related to poverty thresholds and welfare benefits. Our method enables comprehensive analysis of heterogeneity in experienced inflation rates, which has been widely documented (e.g., \citeauthor{kaplan2017}, \citeyear{kaplan2017}; \citeauthor{jaravel2019}, \citeyear{jaravel2019}; \citeauthor{argente2021}, \citeyear{argente2021}).

The core conceptual insight of our approach is that the Slutsky equation, combined with Shephard's lemma, establishes a relationship between moments of demand and the curvature of the expenditure function. By exploiting this connection, we identify local approximations of money-metric welfare measures that are robust to unrestricted preference heterogeneity. We show that our method provides the best welfare point estimates that can be obtained in cross-sectional data under small price changes.\footnote{In this respect, our result nuances the nonidentification result of \citet{hausman2016individual}, who show that average welfare is not identified from cross-sectional data under arbitrary price changes.} In fact, the entire distribution of welfare effects is point-identified in this setting. Moreover, our approach remains computationally feasible in settings with many goods, thus providing a more comprehensive assessment of the impacts of price changes across different consumer segments.\footnote{\cite{hausman2016individual} derive bounds on average welfare effects in a setting with two goods, while \citet{chernozhukovetal} estimate welfare effects in a multigood setting but rely on panel data.}

To understand how our key idea applies in the two-good case, consider a population of heterogeneous consumers indexed by $\omega$, with uncompensated demand $q^\omega$, compensated demand $h^\omega$, expenditure function $e^\omega$, price $p$, and income $y$. For a small price change $\Delta p := p_1 - p_0$ and baseline utility level $v_0^\omega$, the average compensating variation can be written as
\begin{equation*}
    \begin{split}
        \mathbb{E}_\omega\left[ CV^\omega(p_0, p_1, y) \right] &\approx \Delta p \times \mathbb{E}_\omega\left[ D_p e^\omega(p_0, v_0^\omega)\right] + \frac{(\Delta p)^2}{2} \times  \mathbb{E}_\omega\left[D_p^2 e^\omega(p_0, v_0^\omega)\right]\\
        &= \underbrace{\Delta p \times \mathbb{E}_\omega\left[  q^\omega(p_0, y)\right]}_{\text{ mechanical effect }} + \underbrace{\frac{(\Delta p)^2}{2} \times  \mathbb{E}_\omega\left[D_p h^\omega(p_0, v_0^\omega)\right]}_{\text{ behavioral effect }},
    \end{split}
\end{equation*}
where the approximation is obtained from a Taylor expansion and the equality follows from Shephard’s lemma. Although the mechanical effect is identified from average demand, the behavioral effect, which reflects substitution behavior, is not directly identified. However, the Slutsky equation yields:
\begin{equation*}
    \underbrace{\mathbb{E}_\omega\left[ D_p h^\omega(p_0,v^\omega_0)\right]}_{\text{average substitution effect}}=\underbrace{\mathbb{E}_\omega\left[D_p q^\omega(p_0,y)\right]}_\text{average price effect}+\underbrace{\mathbb{E}_\omega\left[q^\omega(p_0,y)D_y q^\omega(p_0,y)\right]}_{\text{average income effect}}.
\end{equation*}
While the average price effect is identified from the price derivative of average demand, the average income effect depends on a nonlinear function of demand. A key insight is that this term can be inferred from heteroscedasticity in demand with respect to income:
\begin{equation*}
    \underbrace{\mathbb{E}_\omega\left[q^\omega(p_0,y)D_y q^\omega(p_0,y)\right]}_{\text{average income effect}} =\underbrace{\frac{1}{2} \times D_y\mathbb{E}_\omega\left[q^\omega(p_0,y)^2\right]}_{\text{heteroscedasticity w.r.t. income}}.
\end{equation*}
This relationship, in turn, enables identification of the average substitution effect and, via Shephard’s lemma, of the average welfare impact through the curvature of the expenditure function.

The generality of our methodology enables several important extensions. First, we assess the biases introduced by standard CLIs that impose homothetic or homogeneous preference restrictions. In particular, we show that our cost-of-living estimate admits a decomposition that quantifies the shortcomings of these traditional measures. Second, the theoretical insights underlying our approach naturally extend beyond consumer welfare analysis. We illustrate this flexibility by incorporating income effects into the estimation of elasticities of taxable income (e.g., \citeauthor{grubersaez02}, \citeyear{grubersaez02}; \citeauthor{brunsziliak16}, \citeyear{brunsziliak16}). Accounting for income effects is empirically relevant, as recent research documents their quantitative importance \citep{golosov2024wealth}. More broadly, our framework applies to a wide range of settings where sufficient statistics are used to evaluate efficiency or welfare costs \citep{kleven2020}.

In our empirical application, we use detailed household-level scanner data from NielsenIQ spanning $2019$ to $2022$ to quantify how the first COVID-19 price shock translated into welfare losses for US grocery shoppers. The data track purchases of fast-moving consumer goods among a large, nationally representative panel of households. For each household, we observe prices paid and quantities purchased, which allow us to construct household-specific price indices across food categories. In addition to detailed purchase data, the dataset includes rich household demographics, such as household income, size, race, and the education level of each household head. These features enable us to estimate the distributional effects of inflation across and within demographic subgroups.

Using our nonparametric method to estimate the CLI, we find that households would have needed over $8\%$ of their monthly food budget in additional compensation to maintain the same utility in December $2021$ as they had in December $2020$. Importantly, the burden of inflation varies substantially across goods. For example, price increases in dry grocery required compensation exceeding $45\%$ of monthly food expenditures, while the corresponding figure for packaged meat was just $6\%$. These differences underscore that the welfare impact of inflation is highly sensitive to the specific categories in which price increases occur.

Next, we find significant heterogeneity in the welfare losses within
and across demographic groups. Across demographic groups, this heterogeneity is driven in part by differences in price increases. However, these differences do not follow a systematic pattern based on observable household characteristics such as race or education. Overall, a typical household’s CLI deviates by approximately $2$ percentage points from the mean. In addition, within groups, households with higher education levels tend to exhibit greater dispersion in CLI. These findings highlight the relevance of unobserved heterogeneity, such as in shopping patterns or store access, in shaping welfare outcomes.

We then compare our CLI to a first-order approximation that holds shares fixed and ignores behavioral responses to price changes such as substitution across goods or income effects. We find that this approximation underestimates the welfare losses by about $4\%$ when we apply it to a composite basket. Notably, this bias is relatively stable across demographic subgroups. However, when examined for specific categories of goods, the bias is far more variable and can be substantial. For instance, the approximation overestimates the welfare loss from price increases in dry grocery and alcohol by approximately $11\%$ and $40\%$, respectively. These large discrepancies highlight that the accuracy of the first-order approximation depends heavily on the category of goods affected and the curvature of demand.

Finally, our decomposition of the behavioral component of the CLI highlights how essential it is to incorporate preference heterogeneity to measure welfare accurately, especially across demographic groups. Models that assume homogeneous preferences, whether or not they incorporate nonhomotheticity, can fail to capture $20\%$ to $40\%$ of the behavioral response to price shocks for some demographic groups. In contrast, the explained share is always within $5\%$ of the full behavioral response once preference heterogeneity is allowed, underscoring that heterogeneity is the primary force behind behavioral adjustments. 


\paragraph{Related literature.}
Our methodology contributes to a long-standing tradition of estimating consumer welfare nonparametrically using cross-sectional data.\footnote{See \citet{bhattacharya24} for a recent and comprehensive review.} A prominent early approach in this literature treats average demand as if generated by a representative consumer and thereby abstracts from preference heterogeneity \citep[e.g.,][]{hausman1981exact, vartia1983efficient}. Along these lines, \citet{hausman1995nonparametric} obtain point estimates via nonparametric regression, while \citet{FosterHahn2000} and \citet{Blundelletal2003} derive conditions under which such estimates provide a first-order approximation of the true welfare changes. Similarly, \citet*{schlee2007measuring} provides sufficient conditions under which the representative agent framework yields upper bounds on welfare.

Motivated by aggregation results from \cite{gorman53}, this approach typically assumes that preferences are homogeneous or homothetic. However, interpreting average demand as that of a representative consumer is valid only under restrictive conditions \citep{jerison1994optimal, lewbel2001demand}, such as when preferences are homothetic or when income effects are negligible. Unfortunately, these restrictions on preferences are often too strong in practice. Notably, the covariance term in \citet{lewbel2001demand} captures the failure of average demand to satisfy integrability conditions due to preference heterogeneity.

Closer to our approach, \citet{banks1996tax} improve on  first-order approximations by incorporating second-order terms. However, they do not link demand moments to the curvature of the expenditure function, a technique that we exploit to characterize welfare impacts under arbitrary preference heterogeneity. \citet*{hoderleinvanhemsm} provide point identification of quantile demand functions in the two-good case by assuming monotonicity in unobserved scalar heterogeneity. While elegant, their identification strategy requires that individuals retain their rank in the distribution of demand across budget sets.
 
More fundamentally, \citet{hausman2016individual} show that average consumer welfare is not point-identified from cross-sectional data under unrestricted heterogeneity and provide worst-case bounds under mild assumptions. Although their results are robust, their approach can lead to wide welfare intervals, particularly when income effects are large. In contrast, our method provides point estimates of local welfare under general heterogeneity, including in settings with multiple goods and unrestricted income effects.

Relatedly, our insight that variation in demand contains a source of identification in cross-sectional data connects to a broader literature. For example, the idea that the variance of demand is informative about average income effects is noted by \cite{hildenbrand83} and \cite{hoderlein2011many}. More generally, \cite{hoderleinmammen} establish that, in nonseparable models, cross-sectional data identify local average structural derivatives, but not transformations thereof. Their result helps clarify that nonidentification of welfare effects stems from the inability to learn higher-order income effects in cross-sectional data.

A different strand of literature relies on revealed preference methods to bound welfare impacts in random utility models. For instance, \citet*{cosaert18} apply the weak axiom of stochastic revealed preference to repeated cross sections. \citet*{deb22} derive and exploit novel revealed preference conditions over prices in repeated cross sections. \citet*{allen2020counterfactual} use revealed preference conditions for the law of demand that are applicable in panel data. While these methods are robust to functional form assumptions, they typically yield set identification.

Our approach also contributes to the literature on CLIs and inflation heterogeneity. Several recent papers examine how nonhomothetic preferences shape inflation across income levels and consumption baskets \citep{fajgelbaum2016measuring, baqaee2022new, jaravel2023measuring}, and \citet{kaplan2017} construct household-specific inflation indices to reflect differential price exposure across groups. We complement this body of work by allowing for unrestricted heterogeneity in preferences and quantifying the bias introduced by assuming homotheticity.

Finally, our framework contributes to the growing literature on sufficient statistic approaches in public finance. In particular, we extend the tools for estimating elasticities and behavioral responses to settings with income effects and heterogeneous preferences, thereby relaxing the quasi-linear utility assumptions common in earlier work \citep[e.g.,][]{grubersaez02, brunsziliak16}. Our methodology also lends itself to the valuation of redistributive policies and government programs, in line with recent unified frameworks as in \citet{hendren2020unified}, \citet{kleven2020}, and \citet{finkelstein2019}.

\paragraph{Outline of the paper.} This paper is organized as follows. Section~\ref{sec:setup} presents the notation and conceptual framework. Section~\ref{sec:compensated} exploits information on the curvature of the expenditure function to approximate moments of consumer welfare. Section~\ref{sec:extensions} contains extensions of our main result to a quantile-based approach, a decomposition of the behavioral response, and elasticities of taxable income. Section~\ref{sec:empirics} details the application and illustrates the usefulness of our method for assessing changes in consumer welfare. 
Finally, Section~\ref{sec:conclusion} concludes. The appendix provides omitted proofs and additional results.

\section{Framework}\label{sec:setup}

Our framework allows for unrestricted unobserved heterogeneity in preferences. For ease of exposition, we suppress observed individual characteristics from the notation in the rest of the paper; our results can be thought of as conditional on such characteristics.

\paragraph{Setup and notation.} Let $\Omega$ be the universe of preference types. Each preference type $\omega \in \Omega$ can be viewed as an individual with preferences over bundles of $(k+1)$ goods $\bq \in \mathcal{Q} \subseteq \mathbb{R}_{++}^{k+1}$, where $\mathcal{Q}$ is compact and convex. We assume that preferences are represented by smooth and strictly quasi-concave utility function $u^\omega : \mathcal{Q} \to \mathbb{R}$ that is infinitely differentiable everywhere.\footnote{This ensures that the demand functions are infinitely differentiable in prices and income.} Furthermore, we denote prices by $\bp \in \mathcal{P}\subset \mathbb{R}_{++}^{k+1}$ and income by $y \in \mathcal{Y} \subset \mathbb{R}_{++}$. We use Euler's notation for differentiation such that $D_{x_1,x_2}^{m,n} f(x_1, x_2) := \frac{\partial^{m+n} f(x_1,x_2)}{\partial x_{1}^m \partial x_{2}^n}$. The derivative of a vector function with respect to another vector is expressed in numerator layout such that $(D_\mathbf{x} \mathbf{f}(\mathbf{x}))_{ij} = D_{x_j} f_i(\mathbf{x})$. A variable in logarithms is denoted with a tilde: e.g., $\widetilde{\bp} := \log(\bp)$ and $\widetilde{y} := \log(y)$. The symbol $\odot$ denotes the element-wise product. Throughout, we use $O$ to denote Landau's big O.\footnote{Formally, $f(x) = O(g(x))$ means there exists constants $C >0$ and $x_0$ such that $|f(x)| \leq C|g(x)|$ for all $x \geq x_0$.}

\paragraph{Consumer demand.} We consider the standard model of utility maximization under a linear budget constraint. Accordingly, an individual demand function $\bq^\omega(\bp, y) : \mathcal{P} \times \mathcal{Y} \to \mathcal{Q}$ is defined as
\begin{equation*}
    \bq^\omega(\bp,y):=\argmax_{\bp^\intercal \mathbf{q}\leq y : \mathbf{q} \in \mathcal{Q}}u^\omega(\bq).
\end{equation*}
For every uncompensated (or Marshallian) demand function $\bq^\omega$, there exists a compensated (or Hicksian) demand function $\bh^\omega(\bp, u) : \mathcal{P} \times \mathbb{R} \to \mathcal{Q}$ defined as 
\begin{equation*}
    \bh^\omega(\bp,u):=\argmin_{\mathbf{q} \in \mathcal{Q}}\{\bp^\intercal \mathbf{\bq} | u^\omega(\bq)\geq u\}.
\end{equation*}
Observe that both demand functions are linked through the Slutsky equation.\footnote{The Slutsky equation is defined as $D_\bp \bh^\omega(\bp, u^\omega(\bq^\omega(\bp, y)))  = D_\bp \bq^\omega(\bp,y)  +  D_y \bq^\omega(\bp,y) \bq^\omega(\bp,y)^\intercal$.} Note also that, without loss of generality, one can omit the demand and price for the $(k+1)$st good because of Walras's law. Next, the indirect utility function $v^\omega(\bp,y) : \mathcal{P} \times \mathcal{Y} \to \mathbb{R}$ is defined as
\begin{equation*}
    v^\omega(\bp,y):=\max_{\bp^\intercal \bq\leq y : \bq \in \mathcal{Q}}u^\omega(\bq)
\end{equation*}
and gives the maximum utility level obtained for the budget set defined by $(\bp,y)$. Likewise, the expenditure function $e^\omega(\bp, u) : \mathcal{P} \times \mathbb{R} \to \mathcal{Y}$ is defined as
\begin{equation*}
    e^\omega(\bp, u):=\min_{u \leq u^\omega(\bq) : \bq \in \mathcal{Q}} \bp^\intercal \bq
\end{equation*}
and gives the minimum amount of income needed to achieve utility level $u$ at prices $\bp$.

\paragraph{Moments of consumer demand.}
To avoid complex tensor notation, it will be convenient to introduce scalar-valued \textit{composite demands}. For any $\bt \in \mathbb{R}^k$, let
\begin{equation*}
    \begin{split}
        q^\omega(\bp,y; \bt) &:= \bt^\intercal \bq^\omega(\bp,y), \\
        h^\omega(\bp, u; \bt) &:= \bt^\intercal \bh^\omega(\bp,u) \\
    \end{split}
\end{equation*}
denote the composite uncompensated and compensated demand, respectively. These composite demands can be interpreted as the projection of demand bundles on the line through the origin defined by $\bt$. For any $n \in \mathbb{N}_{++}$, we define the $n$th (noncentral) conditional moment of composite (uncompensated) demand as
\begin{equation}\label{eq:condmom}
    \begin{split}
        M_n(\bp,y; \bt)&:= \mathbb{E}\left[ q^\omega(\bp,y; \bt)^n \mid \bp, y\right] = \int q^\omega(\bp,y; \bt)^n d F(\omega),\\
    \end{split}
\end{equation}
where $F(\omega)$ denotes the distribution of preference types. In particular, note that we recover moments of (noncomposite) demand by setting $\bt = (0,\dots,0,1,0,\dots,0)^\intercal$.  

These moments generalize the concept of the average structural function \citep{blundell_powell_2003}. Under budget set exogeneity, they are nonparametrically identified from cross-sectional data, as they are conditional expectation functions. Specifically, budget set exogeneity implies that $F(\omega \mid \bp,y) = F(\omega)$. 
While the assumption of exogenous budget sets is strong, it is also standard in the literature on nonparametric identification (e.g., see \citeauthor{hausman2016individual}, \citeyear{hausman2016individual}; \citeauthor*{blomquistnewey}, \citeyear{blomquistnewey}). To our knowledge, existing theoretical results do not establish identification under unrestricted preference heterogeneity when budget sets are endogenous in cross-sectional settings. However, certain forms of endogeneity can be addressed by means of a control function approach. We implement this approach to account for endogenous expenditure in our empirical application in Section~\ref{sec:empirics}.

In the remainder of the paper, we assume that every moment exists and is finite. Unless we state otherwise, expectations and moments are always conditional on prices and income. Nevertheless, we refer to such conditional moments as ``moments" for the sake of brevity. Finally, we assume that conditions for the dominated convergence theorem hold such that derivative and integral operators can be interchanged.\footnote{That is, we assume that there exists a function $g : \Omega \rightarrow \mathbb{R}$ such that, for all $(\bp,y) \in \mathcal{P} \times \mathcal{Y}$ and $m,n \in \mathbb{N}$, it holds that $\Big\Vert \myvec\left(D_{\bp, y}^{m,n} \bq^\omega(\bp,y)\right)\Big\Vert \leq g(\omega)$ with $\int g(\omega) dF(\omega) < \infty$.}

\paragraph{Budget shares.}
Some of our results are more conveniently expressed in terms of budget shares. Therefore, define the uncompensated and compensated budget shares as
\begin{equation*}
    \begin{split}
        \mathbf{w}_q^\omega(\bp, y) &:= \frac{\bp \odot \bq^\omega(\bp, y)}{y}, \\
        \mathbf{w}_h^\omega(\bp, u) &:= \frac{\bp \odot \bh^\omega(\bp, u)}{y}
    \end{split}
\end{equation*}
and their associated composite shares as 
\begin{equation*}
    \begin{split}
        w_q^\omega(\bp, y; \bt) &:= \bt^\intercal \mathbf{w}_q^\omega(\bp, y), \\
        w_h^\omega(\bp, u; \bt) &:= \bt^\intercal \mathbf{w}_h^\omega(\bp, u).
    \end{split}
\end{equation*}
For instance, notice that $w_q^\omega(\bp, y; \widetilde{\bp})$ corresponds to Stone's price index \citep{stone1954}. Similarly to the levels of demand, the $n$th moment of the composite compensated budget share is denoted with $W_n(\bp, y; \bt) := \mathbb{E}[w_q^\omega(\bp, y; \bt)^n]$.

\section{Identification}\label{sec:compensated}

In this section, we show that moments of consumer welfare can be approximated locally by observed moments of demand.\footnote{A more in-depth analysis of the informational content of the moments of demand is provided in \citet{maesmalhotra}.} Our identification argument is constructive and naturally leads to plug-in estimators that are simple to implement. A summary of our main argument is presented in Figure~\ref{fig:flowshart0}.


\tikzset{every shadow/.style={fill=none,shadow xshift=0pt,shadow yshift=0pt}}

\begin{figure}[h!]
\centering
\scalebox{0.8}{
\smartdiagramset{module y sep=2.5cm, back arrow disabled=true,module minimum width=4cm,text width =14cm,uniform color list=gray!40 for 4 items
}
\smartdiagram[flow diagram:vertical]{Moments of uncompensated demand, Features of income effects, Features of the curvature of the expenditure function, Approximation to moments of changes in consumer welfare}
\begin{tikzpicture}[overlay]
 \path (module1) -- (module2) node[midway,below=0.05cm, left=0.5cm,draw=none]{\textcolor{black}{Lemma \ref{lem:income}}};
 \path (module2) -- (module3) node[midway,below=0.1cm, left=0.5cm,draw=none]{\textcolor{black}{Proposition~\ref{thm:linearcompe}}} node[midway,below=0.1cm, right=0.5cm,draw=none]{\textcolor{black}{Shephard's lemma \& Slutsky equation}};
  \path (module3) -- (module4) node[midway,below=0.15cm, left=0.5cm,draw=none]{\textcolor{black}{Theorem~\ref{thm:linearallmoments}}} node[midway,below=0.15cm, right=0.5cm,draw=none]{\textcolor{black}{}};

\end{tikzpicture}
}
      \caption{Schematic overview of our main argument}
    \label{fig:flowshart0}
\end{figure}

\subsection{Curvature of the expenditure function}\label{subsec:compderiv}
Before proceeding to our first main result, we establish a fundamental connection between price and income effects and moments of uncompensated demand.
\begin{lem}\label{lem:income}
    For every $n \in \mathbb{N}_{++}$ and $\bt \in \mathbb{R}^k$, it holds that 
    \begin{equation*}
        \begin{split}
            \mathbb{E}\left[q^\omega(\bp,y; \bt)^{n-1}  D_y q^\omega(\bp,y; \bt)\right] &= \frac{1}{n}  D_y M_{n}(\bp,y; \bt), \\
            \mathbb{E}\left[q^\omega(\bp,y; \bt)^{n-1}  D_\bp q^\omega(\bp,y; \bt)\right] &= \frac{1}{n}  D_\bp M_{n}(\bp,y; \bt),
        \end{split}
    \end{equation*}
    at all $(\bp, y) \in \mathcal{P}\times \mathcal{Y}$.
\end{lem}

\begin{proof}
From the definition of moments of demand, we have 
\begin{equation*}
    \begin{split}
        D_y M_{n}(\bp,y; \bt) &= D_y \left(\int q^\omega(\bp,y; \bt)^n dF(\omega)\right) \\
        &= \int  D_y q^\omega(\bp,y; \bt)^n
    dF(\omega) \\ 
    &= n~\mathbb{E}\left[q^\omega(\bp,y; \bt)^{n-1} D_y q^\omega(\bp,y; \bt)\right],
    \end{split}
\end{equation*}
where the second equality follows from the dominated convergence theorem and the third follows from the chain rule. The proof for the price effects is analogous, mutatis mutandis.
\end{proof}

While straightforward, Lemma~\ref{lem:income} serves as the key insight underlying all subsequent results. Indeed, Lemma~\ref{lem:income} establishes that cross-sectional data are informative about the (noncentered) covariance between powers of composite demand and the marginal propensity to consume. The information on these covariances will be crucial when we reconstruct the compensated price responses and conduct welfare analysis. For instance, it implies that the expected income effect is identified from the derivative of the second moment of demand with respect to income:
\begin{equation*}
    \mathbb{E}\left[q^\omega(\bp,y; \bt)  D_y q^\omega(\bp,y; \bt)\right] = \frac{1}{2} D_y M_2(\bp, y; \bt).
\end{equation*}
That the variance of demand is informative about average income effects has been observed before by \citet{hildenbrand83} and \citet{hoderlein2011many}. Lemma~\ref{lem:income} generalizes this key insight to higher-order income effects.

Let the curvatures of the expenditure function and log expenditure function in the direction $\bt$ be defined as
\begin{equation*}
    \begin{split}
        c^\omega(\bp, y; \bt) &:= \bt^\intercal D_\bp^2 e^\omega(\bp, v^\omega(\bp, y)) \bt, \\
        \widetilde{c}^\omega(\bp, y; \bt) &
        := \bt^\intercal D_{\widetilde{\bp}}^2  \widetilde{e}^\omega(\bp, v^\omega(\bp, y)) \bt,
    \end{split}
\end{equation*}

\noindent
respectively. This curvature captures the sensitivity of the consumer's cost of maintaining her current utility to price changes in the direction $\bt$.  Its magnitude plays a central role in the approximation of money-metric welfare measures such as the compensating variation and the CLI (see Section~\ref{sec:welfaremoments}). Using Lemma~\ref{lem:income} and leveraging the Slutsky equation, we can now identify features of the curvature of the expenditure function. 
\begin{prop}\label{thm:linearcompe}
    For every $n \in \mathbb{N}_{++}$ and $\bt \in \mathbb{R}^k$, it holds that 
    \begin{equation*}
        \begin{split}
            \mathbb{E} \left[q^\omega(\bp, y; \bt)^{n-1} c^\omega(\bp, y; \bt) \right] &= \frac{1}{n} D_\bp M_n(\bp, y; \bt)\bt + \frac{1}{n+1} D_y M_{n+1}(\bp, y; \bt), \\
            \mathbb{E} \left[ w^\omega_q(\bp, y; \bt)^{n-1}\widetilde{c}^\omega(\bp, y; \bt) \right] &= \frac{1}{n} D_{\widetilde{\bp}} W_n(\bp, y; \bt)\bt + \frac{1}{n+1} D_{\widetilde{y}} W_{n+1}(\bp, y; \bt) + W_{n+1}(\bp, y; \bt),
        \end{split}
    \end{equation*}
    at all $(\bp, y) \in \mathcal{P}\times \mathcal{Y}$.
\end{prop}
\begin{proof}
    For clarity, we focus on the curvature of the expenditure function; see Appendix~\ref{app:proofs} for the proof for the log expenditure function. For any $(\bp, y) \in \mathcal{P}\times \mathcal{Y}$, we have
    \begin{equation*}
        \begin{split}
            c^\omega(\bp, y; \bt) &= \bt^\intercal D_\bp^2 e^\omega(\bp, v^\omega(\bp, y)) \bt \\
            &= \bt^\intercal D_\bp \bh^\omega(\bp, v^\omega(\bp, y)) \bt \\
            &= \bt^\intercal \left[D_\bp \bq^\omega(\bp, y) + D_y \bq^\omega(\bp,y) \bq^\omega(\bp,y)^\intercal  \right]\bt \\
            &= D_\bp(\bt^\intercal \bq^\omega(\bp, y))\bt + (\bt^\intercal \bq^\omega(\bp, y))D_y(\bt^\intercal \bq^\omega(\bp, y)) \\
            &= D_\bp q^\omega(\bp, y; \bt)\bt + q^\omega(\bp, y; \bt)D_y q^\omega(\bp, y; \bt),
        \end{split}
    \end{equation*}
    where the second equality follows from Shephard's lemma and the third from the Slutsky equation. Therefore,
    \begin{equation*}
        \begin{split}
            \mathbb{E} \left[q^\omega(\bp, y; \bt)^{n-1} c^\omega(\bp, y; \bt) \right] 
            &= \mathbb{E} [q^\omega(\bp, y; \bt)^{n-1}[D_\bp q^\omega(\bp, y; \bt)\bt + q^\omega(\bp, y; \bt)D_y q^\omega(\bp, y; \bt)]] \\
            &= \frac{1}{n} D_\bp M_n(\bp, y; \bt)\bt + \frac{1}{n+1} D_y M_{n+1}(\bp, y; \bt),
        \end{split}
    \end{equation*}
     where the last equality follows from Lemma~\ref{lem:income}.
\end{proof}

\begin{rem}
    Proposition \ref{thm:linearcompe} implies that moments of demand contain information about compensated price responses. Indeed, by Shephard's lemma, we have
    \begin{equation*}
        \begin{split}
            c^\omega(\bp, y; \bt) =  \bt^\intercal D_\bp \bh^\omega(\bp, v^\omega(\bp, y)) \bt.
        \end{split}
    \end{equation*}
    For instance, Proposition~\ref{thm:linearcompe} implies that the average substitution matrix $\mathbb{E}[D_\bp \bh^\omega(\bp, v^\omega(\bp,y)]$ is identified from cross-sectional data since a symmetric matrix is uniquely determined by its quadratic forms. Using arguments analogous to those above, one can show that
    \begin{equation*}
    \begin{split}
        \mathbb{E}[D_\bp \bh^\omega(\bp, v^\omega(\bp,y)] & =\frac{1}{2}\left[D_\bp \bM_1(\bp,y)+ \left(D_\bp\bM_1(\bp,y)\right)^\intercal+ D_y \bM_2(\bp,y)\right],
    \end{split}
    \end{equation*}
    where $\bM_1(\bp, y)$ is the vector of first moments and $\bM_2(\bp, y)$ the matrix of second moments.\footnote{This result follows, for example, by adding the Slutsky equation to its transpose and taking expectations.} Since every term on the right-hand side is nonparametrically identified from cross-sectional data, the average price response is itself identified.
\end{rem}

\begin{rem}
    In the often-studied two-good case, one can obtain a first-order approximation of the $n$th moment of compensated demand at a counterfactual price $p'$ through the $n$th and $(n+1)$st moments of  demand. Using a first-order Taylor approximation around $p' \approx p$, we have that
    \begin{equation*}
    \begin{split}
        h^\omega(p', v^\omega(p,y))^n &=  h^\omega(p, v^\omega(p,y))^n + \left[h^\omega(p, v^\omega(p,y))^{n-1} D_p h^\omega(p, v^\omega(p,y)) \right]  (p'-p) + O((p'-p)^2)\\
        &=  q^\omega(p, y)^n + \left[q^\omega(p, y)^{n-1} D_p h^\omega(p, v^\omega(p,y)) \right]  (p'-p) + O((p'-p)^2)\\
    \end{split}
\end{equation*}
    Taking expectations on both sides and using Lemma~\ref{lem:income} gives
    \begin{equation*} 
        \begin{split}
            \mathbb{E}[h^\omega(p', v^\omega(p,y))^n] = M_n(p, y) + \left(\frac{1}{n} D_p M_n(p,y) + \frac{1}{n+1} D_y M_{n+1}(p,y)\right) (p' - p) + O((p' - p)^2).
        \end{split}
    \end{equation*}
\end{rem}

\subsection{Consumer welfare}\label{sec:welfaremoments}
Since individual preferences are unobservable, consumer welfare is inherently stochastic from the perspective of the analyst. In this section, we examine how the distribution of consumer welfare can be identified and recovered from cross-sectional data. 

\paragraph{Measures of consumer welfare.}

We derive results for two widely used money-metric measures of consumer welfare: the \emph{compensating variation} (CV) and the \emph{cost-of-living index} (CLI). Consider a (potentially counterfactual) price change from $\bp_0$ to $\bp_1$ and define the absolute and logarithmic price differences as $\Delta \bp := \bp_1 - \bp_0$ and $\widetilde{\Delta \bp} := \widetilde{\bp}_1 - \widetilde{\bp}_0$, respectively. Further define the indirect utilities as $v^\omega_0 := v^\omega(\bp_0, y)$ and $v^\omega_1 := v^\omega(\bp_1, y)$, respectively.

The CV represents the amount of income an individual would require after a price change to restore her initial utility level. Formally, it is defined as\footnote{For clarity of exposition, we adopt a sign convention for the CV and CLI that differs from the standard textbook definition (e.g., see \citeauthor*{mascolelletal95}, \citeyear{mascolelletal95}).}
\begin{equation}\label{eq:CV}
    \begin{split}
        CV^\omega(\bp_0, \bp_1, y) &:=  e^\omega(\bp_1, v^\omega_0) - e^\omega(\bp_1, v^\omega_1) \\
        &= e^\omega(\bp_1, v^\omega_0) - y.
    \end{split}
\end{equation}
Likewise, the CLI represents the proportional change in income an individual would require after a price change to restore her initial utility level. Formally, it is defined as 
\begin{equation}\label{eq:COLI}
    \begin{split}
        CLI^\omega({\bp}_0, {\bp}_1, y) &:= \log \frac{e^\omega({\bp}_1, v^\omega_0)}{e^\omega({\bp}_1, v^\omega_1)} \\
        &= \widetilde{e}^\omega({\bp}_1, v^\omega_0) - \widetilde{y}.
    \end{split}
\end{equation}

Note that both $CV^\omega(\bp_0,\bp_1, y)$ and $CLI^\omega(\bp_0,\bp_1, y)$ are positive whenever $\bp_1 > \bp_0$, reflecting a loss in purchasing power due to a price increase. Conversely, both measures are negative when $\bp_1 < \bp_0$. We assume uniformly bounded income effects to ensure that consumer welfare is bounded (see Appendix~\ref{app:boundedness}).

\paragraph{Identification of local effects.} 
Building on our earlier result on the curvature of the expenditure function, we now state the central result. The theorem below demonstrates that moments of consumer welfare can be approximated in terms of moments of demand.\footnote{Since consumer welfare is bounded, standard results on the Hausdorff moment problem ensure that the distributions of the CV and CLI are uniquely determined by their moments.} The identification argument is constructive, providing a basis for straightforward plug-in estimators.
\begin{thm}\label{thm:linearallmoments}
    The $(n+1)$st-order approximation of the $n$th moment of the CV and CLI depends \textit{only} on the $n$th and $(n+1)$st moment of demand:
    \begin{equation*}
        \begin{split}
            \mathbb{E}[CV^\omega(\bp_0, \bp_1, y)^n] &= \mathfrak{M}_n^{CV}(\bp_0, y; \Delta \bp) + \mathfrak{B}_n^{CV}(\bp_0, y; \Delta \bp) + O(||\Delta \bp||^{n+2}),  \\
            \mathbb{E}[CLI^\omega(\bp_0, \bp_1, y)^n] &= \mathfrak{M}_n^{CLI}(\bp_0, y; \widetilde{\Delta \bp}) +  \mathfrak{B}_n^{CLI}(\bp_0, y; \widetilde{\Delta \bp}) + O(||\widetilde{\Delta \bp}||^{n+2}),
        \end{split}
   \end{equation*}
    where
    \begin{equation*}
        \begin{split}
            \mathfrak{M}_n^{CV}(\bp_0, y; \Delta \bp) &= M_n(\bp_0, y; \Delta \bp), \\
            \mathfrak{B}_n^{CV}(\bp_0, y; \Delta \bp) &= \frac{1}{2} \left(D_\bp M_n(\bp_0, y; \Delta \bp)\Delta \bp + \frac{n}{n+1} D_y M_{n+1}(\bp_0, y; \Delta \bp)\right), \\
            \mathfrak{M}_n^{CLI}(\bp_0, y; \widetilde{\Delta \bp}) &=  W_n(\bp_0, y; \widetilde{\Delta \bp}), \\
            \mathfrak{B}_n^{CLI}(\bp_0, y; \widetilde{\Delta \bp}) &= \frac{1}{2} \left(D_{\widetilde{\bp}} W_n(\bp_0, y; \widetilde{\Delta \bp})\widetilde{\Delta \bp} + \frac{n}{n+1} D_{\widetilde{y}} W_{n+1}(\bp_0, y; \widetilde{\Delta \bp}) + nW_{n+1}(\bp_0, y; \widetilde{\Delta \bp})\right).
        \end{split}
    \end{equation*}
\end{thm}
\begin{proof}
    For the sake of brevity, we present only the proof for the CV herein; see Appendix~\ref{app:proofs} for the proof for the CLI. The second-order expansion of the expenditure function for $\Delta \bp \approx \mathbf{0}$ can be written as
    \begin{equation*}
        \begin{split}
            e^\omega(\bp_1, v_0^\omega) &= e^\omega(\bp_0, v_0^\omega) + D_\bp e^\omega(\bp_0, v_0^\omega) \Delta \bp + \frac{1}{2}(\Delta \bp)^\intercal D_\bp^2 e^\omega(\bp_0, v_0^\omega) \Delta \bp + O(||\Delta \bp||^3) \\
            &= y + q^\omega(\bp_0, y; \Delta \bp) + \frac{1}{2}c^\omega(\bp_0, y; \Delta \bp) + O(||\Delta \bp||^3).
        \end{split}
    \end{equation*}
    Therefore, using Equation~\eqref{eq:CV}, the second-order approximation of the CV becomes
    \begin{equation*}
        CV^\omega(\bp_0, \bp_1, y) = q^\omega(\bp_0, y; \Delta \bp) + \frac{1}{2}c^\omega(\bp_0, y; \Delta \bp) + O(||\Delta \bp||^3),
    \end{equation*}
    such that, for higher powers, we obtain
    \begin{equation*}
        CV^\omega(\bp_0, \bp_1, y)^n = q^\omega(\bp_0, y; \Delta \bp)^n + \frac{n}{2} q^\omega(\bp_0, y; \Delta \bp)^{n-1}c^\omega(\bp_0, y; \Delta \bp) + O(||\Delta \bp||^{n+2}).
    \end{equation*}
    Taking expectations on both sides and using Proposition~\ref{thm:linearcompe} gives
    \begin{equation*}
        \begin{split}
            \mathbb{E}[CV^\omega(\bp_0, \bp_1, y)^n] &= \mathfrak{M}_n^{CV}(\bp_0, y; \Delta \bp) + \mathfrak{B}_n^{CV}(\bp_0, y; \Delta \bp)+ O(||\Delta \bp||^{n+2}),
        \end{split}
    \end{equation*}
    \noindent
    as desired.
\end{proof}
In these expressions, the terms  $\mathfrak{M}_n^{CV}(\bp_0, y; \Delta \bp)$ and $\mathfrak{M}_n^{CLI}(\bp_0, y; \widetilde{\Delta \bp})$ represent the mechanical contributions of price changes to the $n$th moments of consumer welfare---specifically, the effects that would arise if consumers did not adjust their behavior. In contrast, the terms $\mathfrak{B}_n^{CV}(\bp_0, y; \Delta \bp)$ and $\mathfrak{B}_n^{CLI}(\bp_0, y; \widetilde{\Delta \bp})$ capture behavioral adjustments, reflecting the reoptimization of consumer choices in response to the change in prices.

\begin{rem}
    Theorem~\ref{thm:linearallmoments} admits further simplification in two empirically relevant cases. First, in the two-good setting, the approximation for moments of the CV simplifies to
    \begin{equation*}
        \mathbb{E}[CV^\omega(p_0, p_1, y)^n] \approx (\Delta p)^n\left(M_n(p_0,y)+\frac{\Delta p}{2}\left[D_p M_n(p_0,y) +\frac{n}{n+  1}  D_y M_{n+1}(p_0,y)\right]\right).
    \end{equation*}
    This case is of practical importance because when the price of a single good changes, only the demand for that good must be modeled. This substantially reduces the dimensionality of the commodity space \citep*{hausman1981exact}. Second, even in the many-good setting, the second-order approximation of the average CV depends solely on the first two moments of demand:
    \begin{equation*}
    \begin{split}
        \mathbb{E}[CV^\omega(\bp_0, \bp_1, y)] \approx M_{1}(\bp, y; \Delta \bp) + \frac{1}{2} \left(D_\bp M_{1}(\bp, y; \Delta \bp) \Delta \bp + \frac{1}{2} D_y M_{2}(\bp, y; \Delta \bp)\right).
    \end{split}
    \end{equation*}
    Thus, average welfare effects can be computed directly from estimates of the mean and variance of demand, offering a tractable approach in applied work. Analogous simplifications hold for the CLI.
\end{rem}

\begin{rem}
A useful way to interpret our approach to estimating the CLI is as a localized application of the {translog unit cost function}  \citep{christensenjorgensonlau, diewert76}. Specifically, the approximation developed in Theorem~\ref{thm:linearallmoments} corresponds to a second-order expansion of the log expenditure function around observed prices. This is similar to evaluating welfare by means of the translog form
\begin{equation*}
    \widetilde{e}_{TL}^\omega(\bp, u) := \alpha_0 + \widetilde{\bp}^\intercal \boldsymbol{\alpha}_1 + \frac{1}{2} \widetilde{\bp}^\intercal \boldsymbol{\Gamma} \widetilde{\bp},
\end{equation*}
where $\mathbf{1}^\intercal \boldsymbol{\alpha}_1 = 1$, $\boldsymbol{\Gamma}$ is symmetric and $\boldsymbol{\Gamma} \mathbf{1} = \mathbf{0}$. Our method can therefore be seen as recovering a local translog representation of preferences to approximate the welfare effects of price changes without imposing the global structure of a fully specified demand system.
\end{rem}

\paragraph{Nonidentification of global effects.}
Our local approximations of consumer welfare represent the most informative estimates obtainable in a cross-sectional setting. The inability to identify higher-order approximations stems from the fact that cross-sectional data lack information about how the variance---and higher moments---of income effects vary across demand bundles. While the issue of nonidentification was previously noted by \citet{hausman2016individual}, Proposition~\ref{thm:nonidentificationhigher} sharpens this insight by formally characterizing the limits of identification.

\begin{prop}\label{thm:nonidentificationhigher}
The $(n+2)$nd-order approximation of the $n$th moment of the CV or CLI is not identified.
\end{prop}
\begin{proof}For clarity, we focus on the average CV with two goods. Suppose the true series expansion of the average CV at a given budget set $(p_0, y)$ is $\mathbb{E}[CV^\omega(p_0,p_1,y)]=a_0 +a_1\Delta p+a_2(\Delta p)^2+a_3(\Delta p)^3$. Extending the argument from the proof of Theorem~\ref{thm:linearallmoments}, recovering $a_3$ requires identifying $\mathbb{E}\left[D_p^2 h^\omega(p_0,v^\omega_0) \right]$, i.e., the expected second derivative of compensated demand with respect to price. Differentiating the identity $h^\omega(p_0,v_0^\omega) \equiv q^\omega(p_0,e^\omega(p_0,v_0^\omega))$ twice with respect to price and substituting in the Slutsky equation, we obtain
\begin{equation*}
    \begin{split}
        D_p^2 h^\omega(p_0,v^\omega_0) &= D_p^2 q^\omega(p_0,y)+ q^\omega(p_0,y) D_{p,y} q^\omega(p_0,y) \\
        &\qquad + D_p q^\omega(p_0,y) D_y q^\omega(p_0,y) + q^\omega(p_0,y) \left(D_y q^\omega(p_0,y)\right)^2 .
    \end{split}
\end{equation*}
Taking expectations and interchanging differentiation and integration yields
\begin{equation*}
    \begin{split}
\mathbb{E}\left[D^2_p h^\omega(p_0, v^\omega_0)\right]  &= D_p^2 M_1(p_0,y)  + \frac{1}{2} D_{p,y} M_2(p_0,y) + \mathbb{E}\left[q^\omega(p_0,y) \left( D_y q^\omega(p_0,y)\right)^2 \right].
    \end{split}
\end{equation*}
As a consequence of Lemma~\ref{lem:nonidentification} in the appendix, the final term on the right-hand side cannot be identified from cross-sectional data: Two observationally equivalent models can yield different values for $\mathbb{E}\left[q^\omega(p_0,y) \left(D_y q^\omega(p_0,y)\right)^2  \right]$. Consequently, the third-order approximation of the average CV is also not identified.
\end{proof}

    The proof of Proposition~\ref{thm:nonidentificationhigher} reveals that the source of nonidentification in cross-sectional data stems from terms such as $\mathbb{E}\left[q^\omega(p_0,y) \left(D_y q^\omega(p_0,y)\right)^2  \right]$. To understand the underlying issue, note that by the law of iterated expectations, this term can be rewritten as
    \begin{equation*}
        \mathbb{E}\left[q^\omega(p_0,y) \left(D_y q^\omega(p_0,y)\right)^2  \right] = \mathbb{E}\left[q^\omega(p_0,y) \mathbb{E}\left[\left( D_y q^\omega(p_0,y)\right)^2 \mid q^\omega(p_0,y)\right]\right].
    \end{equation*}
    The right-hand side represents the (noncentered) covariance between the demand bundle and the second moment of the income effect at that demand bundle. Consequently, the failure to identify the third-order approximation of the average CV arises because cross-sectional data do not provide information on how the variance of the income effect varies across demand bundles. 

    This negative finding is closely linked to fundamental results in the literature on nonseparable models. A direct application of Theorem 2.1 in \citet*{hoderleinmammen} establishes that, in nonseparable models, cross-sectional data identify local average structural derivatives (e.g., $\mathbb{E}\left[ D_y q^\omega(p_0,y) \mid q^\omega(p_0,y)\right]$), but not transformations of these derivatives (e.g., $\mathbb{E}\left[\left( D_y q^\omega(p_0,y)\right)^2 \mid q^\omega(p_0,y)\right]$). As a result, while moments of compensated price derivatives are identified, moments of second-order price derivatives are not. This reasoning extends naturally to higher-order approximations and the many-good case, mutatis mutandis.


\section{Further results and extensions}\label{sec:extensions}
The results developed in Section~\ref{sec:compensated} extend beyond the direct analysis of moments of consumer welfare. In this section, we briefly explore extensions and alternative applications, including decomposition of CLIs and estimation of taxable income elasticities.

\subsection{A quantile-based estimator of consumer welfare}\label{sec:quantile}
If the objective is to recover the full distribution of consumer welfare, an alternative strategy based on quantile regression can be employed. This approach leverages information in the quantiles of  composite demand.\footnote{\citet{Dettehoderlein2016} and \citet{gunsilius2025nonparametrictestslutskysymmetry} leverage quantiles of linear combinations of demand to test for consumer rationality.}

Let $F_q(z \mid \bp, y; \bt) := \Pr[ q^\omega(\bp, y; \bt)  \leq z \mid \bp, y]$ denote the conditional cumulative distribution function (CDF) of composite demand, and let the corresponding conditional quantile function be defined by $K_{q, \tau}(\bp, y; \bt) := \inf \{z : \tau \leq F_q(z \mid \bp, y; \bt)\}$ for all $\tau \in (0,1)$. Similarly, for budget shares, define the conditional CDF by $F_w(z \mid \bp, y; \bt) := \Pr[ w_q^\omega(\bp, y; \bt)  \leq z \mid \bp, y]$ and the associated conditional quantile function by $K_{w, \tau}(\bp, y; \bt) := \inf \{z : \tau \leq F_w(z \mid \bp, y; \bt)\}$.
\begin{prop}\label{prop:quantile}
    Under standard regularity conditions \citep{hoderleinmammen}, the CDFs of the CV and CLI can be approximated as
    \begin{equation*}
        \begin{split}
            F_{CV}(z \mid \bp_0, y; \Delta \bp) &\approx \Pr_{\tau}\left[K_{q, \tau}(\bp_0, y ; \Delta \bp) + S_\tau^{CV}(\bp_0, y; \Delta \bp) \leq z \mid \bp_0, y;  \Delta \bp\right], \\
            F_{CLI}(z \mid \bp_0, y; \widetilde{\Delta \bp}) &\approx \Pr_{\tau}\left[K_{w, \tau}(\bp_0, y ; \widetilde{\Delta \bp}) + S_\tau^{CLI}(\bp_0, y; \widetilde{\Delta \bp}) \leq z \mid \bp_0, y; \widetilde{\Delta \bp}\right],
        \end{split}
    \end{equation*}
    where $\tau \sim U(0,1)$ and 
    \begin{equation*}
        \begin{split}
            S_\tau^{CV}(\bp_0, y; \Delta \bp) &= \frac{1}{2}  \left(D_\bp K_{q, \tau}(\bp_0, y ; \Delta \bp)  \Delta \bp +  K_{q, \tau}(\bp_0, y ; \Delta \bp) D_y K_{q, \tau}(\bp_0, y ; \Delta \bp) \right), \\
            S_\tau^{CLI}(\bp_0, y; \widetilde{\Delta \bp}) &= \frac{1}{2} \left(D_{\widetilde{\bp}}K_{w, \tau}(\bp_0, y ; \widetilde{\Delta \bp})\widetilde{\Delta \bp} + K_{w, \tau}(\bp_0, y ; \widetilde{\Delta \bp})D_{\widetilde{y}}K_{w, \tau}(\bp_0, y ; \widetilde{\Delta \bp}) + K_{w, \tau}(\bp_0, y ; \widetilde{\Delta \bp})^2\right). \\
        \end{split}
    \end{equation*}
\end{prop}
\begin{proof}
    See Appendix~\ref{app:proofs}.
\end{proof}

Proposition~\ref{prop:quantile} may be viewed as a local extension of the result established by \citet{hoderleinvanhemsm}, generalized to accommodate multiple goods and unrestricted unobserved heterogeneity. In the two-good case, \citet{hoderleinvanhemsm} show using conditional quantile demands that exact individual-level welfare analysis is feasible under the restrictive assumption that demand is monotonic in scalar-valued unobserved heterogeneity.\footnote{Under this assumption, the conditional quantiles coincide with individual demand functions, and consumer welfare can be computed by the method of \citet{vartia1983efficient}.} Proposition~\ref{prop:quantile} demonstrates that a version of this insight carries over to the more general setting with multiple goods and arbitrary heterogeneity provided that the price changes are sufficiently small. In this local setting, the distribution of consumer welfare can be approximated by means of the average compensated responses of hypothetical consumers located at different quantiles of the composite demand distribution. In particular, we show that all moments of these approximate welfare distributions match those derived in Theorem~\ref{thm:linearallmoments}, up to the same order of approximation. Whether more accurate quantile-based approximations exist remains an open question, which is left for future research.

\subsection{Decomposition of CLIs}
A growing literature raises the point that standard price index formulas suffer from a bias when preferences are nonhomothetic \citep{baqaee2022new,jaravel2023measuring,fajgelbaum2016measuring}.
Theorem~\ref{thm:linearallmoments} allows us to decompose the bias of standard formulas into the components arising from heterogeneity and nonhomotheticity in preferences.\footnote{Interestingly, in our empirical application in Section~\ref{sec:empirics}, we find that the contribution to the bias from nonhomotheticity is more substantial than that from heterogeneity.} Such a decomposition can be used to assess the relative importance of accounting for the two features. 
\begin{prop}\label{cor:decomposition}
    The behavioral term of the aggregate price index can be decomposed as 
    \begin{equation*}
    \begin{split}
        \mathfrak{B}_n^{CLI}(\bp_0, y; \widetilde{\Delta \bp}) =
        \sum_{k=1}^4 \mathfrak{D}_{n, k}(\bp_0, y; \widetilde{\Delta \bp}),
    \end{split}
\end{equation*}
where
    \begin{equation*}
       \begin{split}
           \mathfrak{D}_{n, 1}(\bp_0, y; \widetilde{\Delta \bp}) &:=  \frac{1}{2} \left(D_{\widetilde{\bp}} W_1(\bp_0, y; \widetilde{\Delta \bp})^n\widetilde{\Delta \bp} + nW_{1}(\bp_0, y; \widetilde{\Delta \bp})^{n+1}\right), \\
           \mathfrak{D}_{n, 2}(\bp_0, y; \widetilde{\Delta \bp}) &:=  \frac{1}{2}\left(\frac{n}{n+1} D_{\widetilde{y}} W_{1}(\bp_0, y; \widetilde{\Delta \bp})^{n+1}\right), \\
           \mathfrak{D}_{n, 3}(\bp_0, y; \widetilde{\Delta \bp}) &:= \frac{1}{2}\left(   D_{\widetilde{\bp}} \overline{W}_n(\bp_0, y; \widetilde{\Delta \bp})\widetilde{\Delta \bp} + n\overline{W}_{n+1}(\bp_0, y; \widetilde{\Delta \bp})\right),\\
           \mathfrak{D}_{n, 4}(\bp_0, y; \widetilde{\Delta \bp}) &:= \frac{1}{2}\left( \frac{n}{n+1} D_{\widetilde{y}} \overline{W}_{n+1}(\bp_0, y; \widetilde{\Delta \bp})\right),\\
       \end{split} 
    \end{equation*}
    and $\overline{W}_n(\bp_0, y; \bt) := {W}_n(\bp_0, y; \bt) - {W}_1(\bp_0, y; \bt)^n$.
\end{prop}
\begin{proof}
    See Appendix~\ref{app:proofs}.
\end{proof}
Table~\ref{tab:deccomposition} provides a schematic overview of our decomposition. The term $\mathfrak{D}_{n, 1}$ captures the behavioral component under the assumption of a homothetic representative agent (RA). Adding $\mathfrak{D}_{n, 2}$ relaxes this assumption by allowing the RA to have nonhomothetic preferences. The sum $\mathfrak{D}_{n, 1} + \mathfrak{D}_{n, 2}$ corresponds to the standard RA approximation. 

The term $\mathfrak{D}_{n, 3}$ captures the effect of preference heterogeneity in a population with homothetic preferences. That is, $\mathfrak{D}_{n, 1} + \mathfrak{D}_{n, 3}$ yields an approximation for a heterogeneous population, assuming homotheticity. Finally, $\mathfrak{D}_{n, 4}$ captures deviations from homotheticity at the population level. The sum $\mathfrak{D}_{n, 3} + \mathfrak{D}_{n, 4}$ represents the deviation between our full approximation and the RA benchmark.

\begin{table}[h]
\centering
\caption{Decomposition of the behavioral term of the CLI}\label{tab:deccomposition}
\begin{tabular}{lll}
\toprule
Preferences            & Homothetic  & Nonhomothetic          \\
\midrule
Homogeneous   & $\mathfrak{D}_{n, 1}$       & $\mathfrak{D}_{n, 1} + \mathfrak{D}_{n, 2}$             \\
Heterogeneous & $\mathfrak{D}_{n, 1} + \mathfrak{D}_{n, 3}$ & $\mathfrak{D}_{n, 1} + \mathfrak{D}_{n, 2} + \mathfrak{D}_{n, 3} + \mathfrak{D}_{n, 4}$ \\
\bottomrule
\end{tabular}
\end{table}
 
Using an RA approach, \citet{jaravel2023measuring} propose a correction for nonhomothetic preferences based on $D_y \mathbb{E}[CLI^\omega(\bp_0, \bp_1, y)]$, which captures the sensitivity of the aggregate CLI to income changes. Our approach extends this analysis by incorporating unobserved preference heterogeneity into the same parameter. Formally, 
\begin{equation*}
    \begin{split}
        D_y \mathbb{E}\left[CLI^\omega(\bp_0, \bp_1, y)\right] &= \underbrace{D_y \left[\mathfrak{M}_1^{CLI}(\bp_0, y; \widetilde{\Delta \bp}) +\mathfrak{D}_{1, 1}(\bp_0, y; \widetilde{\Delta \bp}) + \mathfrak{D}_{1, 2}(\bp_0, y; \widetilde{\Delta \bp})\right]}_{\text{correction for preference nonhomotheticity}} \\
        &\quad + \underbrace{D_y \left[\mathfrak{D}_{1, 3}(\bp_0, y; \widetilde{\Delta \bp}) + \mathfrak{D}_{1, 4}(\bp_0, y; \widetilde{\Delta \bp})\right]}_{\text{correction for preference heterogeneity}}.
    \end{split}
\end{equation*}
Our decomposition allows direct assessment of the relative contributions of nonhomotheticity and heterogeneity in preferences to the overall income sensitivity of the aggregate CLI.

\subsection{Elasticity of taxable income}
A central theme in the public economics literature concerns the measurement of the impact of income tax reforms. If a reform is small, a sufficient-statistic approach based on compensated elasticities delivers a convenient approximation of the welfare cost of taxation \citep{feldstein99, grubersaez02, chetty2009}.\footnote{Income effects enter the first-order approximations because of the nonlinear nature of the tax schedule. That is, virtual income (see infra) is affected by price or tax changes.} In practice, however, most studies proceed by means of uncompensated elasticities, implicitly assuming that individuals have quasi-linear utilities (e.g., see \cite{brunsziliak16} and the references therein). When this assumption is unwarranted, this can bias the efficiency estimates. Our method provides a means of accounting for income effects without increasing data requirements.

Consider a setting with two goods: before-tax income $q^\omega$ and consumption $c^\omega$. An individual derives utility from consumption and disutility from before-tax income, as the latter requires effort to earn. To allow for nonproportional taxation, we assume the tax schedule is piecewise linear. At a linear part of this schedule, the budget constraint is given by $c^\omega = (1-\tau)q^\omega + r$, in which $\tau$ denotes the (local) marginal tax rate and $r$ virtual income.\footnote{Following the literature, we abstract from individuals located at the kinks of the budget set.} Figure~\ref{fig:piecewise} illustrates this relationship, showing the piecewise-linear budget constraint alongside its linear approximation at a specific segment (indicated by the dashed line).

\begin{figure}[h!]
  \centering
\begin{tikzpicture}[scale=1]
  \draw[->] (0,0) -- (8,0) node[right] {\(q^\omega\)};
  \draw[->] (0,0) -- (0,5) node[above] {\(c^\omega\)};

  \def\qOne{2}      
  \def\qTwo{5}      
  \def\slopeOne{0.9}    
  \def\slopeTwo{0.5}    
  \def\slopeThree{0.3}  

  \pgfmathsetmacro{\cOne}{\slopeOne * \qOne}
  \pgfmathsetmacro{\cTwo}{\cOne + \slopeTwo * (\qTwo - \qOne)}

  \coordinate (O) at (0,0);
  \coordinate (A) at (\qOne, \cOne);
  \coordinate (B) at (\qTwo, \cTwo);
  \coordinate (I) at (0, 0.8);
  \coordinate (E) at (8, 4.8);

  \draw[thick, black] (O) -- (A);
  \draw[thick, black] (A) -- (B);
  \draw[thick, black] (B) -- ++(2.5, {\slopeThree * 2.5});

  \draw[black, dashed] (I) -- (E);

  \def\a{0.0625}
  \def\U{1.8}
  \draw[thick, black, domain=1.5:6.5, smooth, variable=\x] 
    plot ({\x}, {\U + \a*(\x)^2});

    \draw[dotted, black] (4.8,2.3) -- (4.8,3.2);
\draw[dotted, black] (3, 2.3) -- (4.8, 2.3);

  \node at (5.5, 2.8) {\(1-\tau\)};
  \node at (3.9, 2) {$1$};
  \node[anchor=east] at (I) {\(r\)};
  \node[black] at (5, 5) {indifference curve};
\node[black] at (8.5, 3.4) {budget constraint};
  \filldraw[black] (4,2.8) circle (1.5pt);
\end{tikzpicture}
  \caption{A piecewise-linear budget constraint}\label{fig:piecewise}
\end{figure}

\citet{feldstein95, feldstein99} shows that the welfare cost of taxation can be summarized by means of the compensated elasticity of taxable income 
\begin{equation*}
    \varepsilon^\omega(\tau, v_0^\omega) := \frac{(1-\tau)}{h^\omega(1-\tau, v_0^\omega)}D_{1-\tau} h^\omega(1-\tau, v_0^\omega), 
\end{equation*}
where $h^\omega$ denotes compensated demand for before-tax income. Insights similar to those in Section~\ref{sec:compensated} allow us to nonparametrically identify and estimate the average of this compensated price elasticity.\footnote{This result is related to the work of \citet*{paluchkneiphildenbrand}, who derive a connection between individual and aggregate income elasticities.}

\begin{prop}
    The average compensated elasticity of taxable income can be written as
    \begin{equation*} 
        \mathbb{E}[\varepsilon^\omega(\tau, v_0^\omega)] = (1-\tau) \Big(D_{1-\tau}\mathbb{E}[\widetilde{q}^\omega(1-\tau,r)] + D_r \mathbb{E}[q^\omega(1-\tau,r)]\Big).
    \end{equation*}
\end{prop}
\begin{proof} Following arguments similar to those above, it holds that
    \begin{equation*}
        \begin{split}
            \mathbb{E}[\varepsilon^\omega(\tau, v_0^\omega)] &= (1-\tau) \mathbb{E}\left[\frac{1}{h^\omega(1-\tau, v_0^\omega)} D_{1-\tau} h^\omega(1-\tau, v_0^\omega) \right] \\
            &= (1-\tau) \mathbb{E}\left[\frac{1}{q^\omega(1-\tau, r)}\Big(D_{1-\tau} q^\omega(1-\tau, r) + q^\omega(1-\tau, r) D_r q^\omega(1-\tau,r)\Big)\right] \\
            &= (1-\tau) \Big(D_{1-\tau} \mathbb{E}[\widetilde{q}^\omega(1-\tau, r)] + D_r \mathbb{E}[q^\omega(1-\tau,r)]\Big).
        \end{split}
    \end{equation*}
\end{proof}

\section{Empirical application: Inflation and consumer welfare}\label{sec:empirics}

This section introduces the data, outlines the estimation, and presents our empirical findings. More precisely, we first apply our results to detailed scanner data on grocery purchases to assess the welfare effects of inflation during the initial COVID-$19$ price shock.\footnote{Our empirical strategy exploits both cross-sectional and intertemporal price variation to identify welfare effects. However, the method is equally applicable in settings with only one source of price variation (e.g., see \citeauthor*{blundellhorowitzparey}, \citeyear{blundellhorowitzparey}).} Then, we compare our CLI estimate against a first-order approximation and show that the latter consistently underestimates the true cost of living. Furthermore, we show there is sizable heterogeneity in the cost of living across households, much of which cannot be explained by observed household characteristics. Finally, we quantify the bias introduced by various restrictions on preferences in the CLI. We find that heterogeneity in preferences is crucial for capturing the true cost of living and that nonhomotheticity further eliminates a meaningful and systematic bias.

\subsection{Sample} \label{sec:appdata}

Our empirical analysis uses household-level scanner data from NielsenIQ, which track fast-moving consumer goods purchased by a representative panel of households in the United States. This dataset covers a wide range of products, including food items and nonfood categories such as cosmetics, pet care, and office supplies. The panel consists of approximately $60$,$000$ households, with some participating for many years and others entering or exiting the panel over time. Using in-home scanners or a mobile application, panelists record the universal product codes (UPCs) for all the products they purchase. For each transaction, we observe the price paid, quantity purchased, and detailed product characteristics. Additionally, the data include rich household-level demographics, such as household income, household size, and the education level of each head household member.

Our sample spans January $2019$ to December $2022$, allowing us to capture price dynamics before, during, and after the peak of the COVID-$19$ lockdown. We construct monthly, household-specific price indices across food departments as defined by NielsenIQ: dry grocery, frozen foods, dairy, deli, packaged meat, fresh produce, and alcohol. Following the approach of \cite{HoderleinMihaleva2008}, these indices are linear in prices, with UPC-level prices weighted by household-specific expenditure shares. This method preserves price heterogeneity and captures substitution patterns as the relative prices faced by each household change. Further details about the price aggregation are provided in Appendix \ref{app:data}.


Our final sample consists of $269$,$593$ household--year--month observations spanning the years $2019$--$2022$. For each observation, we observe prices and expenditures across food departments. We equivalize household income and total expenditures using a modified OECD equivalence scale and remove observations with missing information or extreme values in shares, prices, or equivalized expenditures.\footnote{Equivalized expenditure is computed as  $y = y/(1 + 0.5\cdot \mathds{1}(\text{nadults} = 2) + 0.3\cdot \text{nchildren})$, where $y$ is household expenditure, $\mathds{1}(\cdot)$ is the indicator function that equals one if there are two adults in the household (nadults $= 2$), and nchildren is the number of children in the household. The computations of equivalized household income are identical.} The resulting dataset includes only household--year--month observations with nonmissing purchases in each food department category. For expositional simplicity, we may refer to departments as ``goods" throughout the paper. Further details on the sample construction are provided in Appendix \ref{app:data}.

\subsection{Summary statistics}

This section documents prices and expenditures across demographic groups and tracks how prices and expenditure shares have evolved over time across departments. First, Table \ref{tab:logprices} reports summary statistics across demographic groups, highlighting several noteworthy consumption patterns. 
\begin{table}[H]
\centering
\caption{Summary statistics by demographic group}
\label{tab:logprices}
\begin{adjustbox}{width=\textwidth}
\begin{threeparttable}

\newcolumntype{C}[1]{>{\centering\arraybackslash}p{#1}} 
\setlength{\tabcolsep}{5pt} 

\begin{tabular}{l
                C{1.7cm} C{1.7cm} 
                C{1.7cm} C{1.7cm} 
                C{1.7cm} C{1.7cm} 
                C{1.7cm} C{1.7cm} 
                C{1cm}}
\toprule
Demographic & 
\multicolumn{2}{c}{Price} &
\multicolumn{2}{c}{Expenditure$^{\mathrm{a}}$} & 
\multicolumn{2}{c}{Number of children} & 
\multicolumn{2}{c}{Age$^{\mathrm{b}}$} & 
$N$ \\
\cmidrule(lr){2-3} \cmidrule(lr){4-5} \cmidrule(lr){6-7} \cmidrule(lr){8-9}
& Estimate & Std Dev & Estimate & Std Dev & Estimate & Std Dev & Estimate & Std Dev & \\
\midrule
\textit{Household size} & & & & & & & & & \\
\quad $1$ & 4.56 &  1.30 & 273.80 & 98.79 & 0.00 & 0.00 & 61.63 & 13.07 & 18,320 \\
\quad $2$ & 4.33 &  1.30 & 384.88 & 138.24 & 0.00 & 0.00 & 62.36 & 12.70 & 45,379 \\
\quad $3+$ & 4.12 & 1.30 & 426.17 & 168.23 & 1.84 & 0.96 & 50.98 & 11.50 & 31,933 \\
\addlinespace
\textit{Annual income}$^{\mathrm{a}}$ & & & & & & & & & \\
\quad $<25$k & 4.32 & 1.33 & 392.98 & 159.08 & 0.83 & 1.16 & 57.00 & 13.02 & 42,690 \\
\quad $25$k–$35$k & 4.13 & 1.27 & 361.37 & 150.27 & 0.65 & 1.12 & 59.84 & 14.46 & 13,754 \\
\quad $35$k–$50$k & 4.22 & 1.27 & 375.61 & 149.54 & 0.69 & 0.92 & 57.74 & 13.63 & 21,366 \\
\quad $50$k–$70$k & 4.45 & 1.28 & 367.28 & 138.43 & 0.00 & 0.00 & 61.85 & 12.95 & 15,075 \\
\quad $>70$k & 4.89 & 1.31 & 284.64 & 101.38 & 0.00 & 0.00 & 59.94 & 12.42 & 2,747 \\
\addlinespace
\textit{Race}$^{\mathrm{c}}$ & & & & & & & & & \\
\quad White & 4.29 & 1.29 & 380.47 & 151.38 & 0.57 & 1.00 & 59.24 & 13.44 & 76,894 \\
\quad Black & 4.28 & 1.32 & 349.09 & 152.83 & 0.72 & 1.12 & 56.65 & 12.87 & 10,890 \\
\quad Asian & 4.81 & 1.48 & 387.53 & 161.39 & 0.94 & 1.14 & 52.37 & 12.54 & 2,935 \\
\quad Other & 4.38 & 1.37 & 385.84 & 157.56 & 0.86 & 1.18 & 53.20 & 13.22 & 4,913 \\
\addlinespace
\textit{Education}$^{\mathrm{d}}$ & & & & & & & & & \\
\quad Low & 4.22 & 1.27 & 371.66 & 150.20 & 0.54 & 1.00 & 60.39 & 12.88 & 40,627 \\
\quad Mid & 4.33 & 1.32 & 380.49 & 154.13 & 0.67 & 1.05 & 56.65 & 13.52 & 35,324 \\
\quad High & 4.44 & 1.35 & 383.64 & 153.99 & 0.67 & 1.04 & 57.55 & 13.99 & 19,681 \\
\bottomrule
\end{tabular}

\begin{tablenotes}
    \item[a] {\footnotesize To account for differences in household composition, we equivalize annual income using a modified OECD scale. The equivalized income (EI) is computed as EI $= I/(1 + 0.5\cdot \mathds{1}(nadults = 2) + 0.3\cdot nchildren)$, where $I$ is household annual income, $\mathds{1}(\cdot)$ is the indicator variable, $nadults$ is the number of adults, and $nchildren$ is the number of children. Expenditure is adjusted similarly.}
    \item[b] {\footnotesize Age refers to the age of the household head. When two heads are present, age refers to that of the male head.}
    \item[c] {\footnotesize Race represents the racial identity of the household as reported by the panelist.}
    \item[d] {\footnotesize Education level refers to the highest level of education attained by any household head. Low = high school diploma or less; Mid = some college or college degree; High = postgraduate degree.}
\end{tablenotes}

\end{threeparttable}
\end{adjustbox}
\end{table}


The results in the table show that log prices are highest among single-person households and decline with household size, consistent with cost savings from bulk purchases. Similarly, average prices increase with household income and education, reflecting preferences for differentiated or higher-quality brands among higher-income and more educated households. 

Interestingly, total expenditure exhibits an inverse pattern with income whereby households in higher income groups spend less on average than lower-income households. This seemingly counterintuitive result is driven primarily by differences in household composition. Indeed, higher-income households tend to be smaller and often do not have children, which lowers their overall consumption levels. This is confirmed by the corresponding number of children, which falls to zero for the two highest income brackets.

Table \ref{tab:logprices} also shows heterogeneity in prices paid across racial groups. For example, Asian households pay the highest average prices, while Black households pay the lowest, potentially reflecting differences in access to retailers, brand preferences, or store loyalty. Finally, note that the standard deviations in prices and expenditures are typically higher among low-income and larger households, indicating greater variability in their shopping behavior and product selection. 

\begin{figure}[H]
    \centering
    \begin{minipage}[b]{0.49\textwidth}
        \centering
        \subcaption*{Panel A: Average log price by department}
        \includegraphics[width=\linewidth]{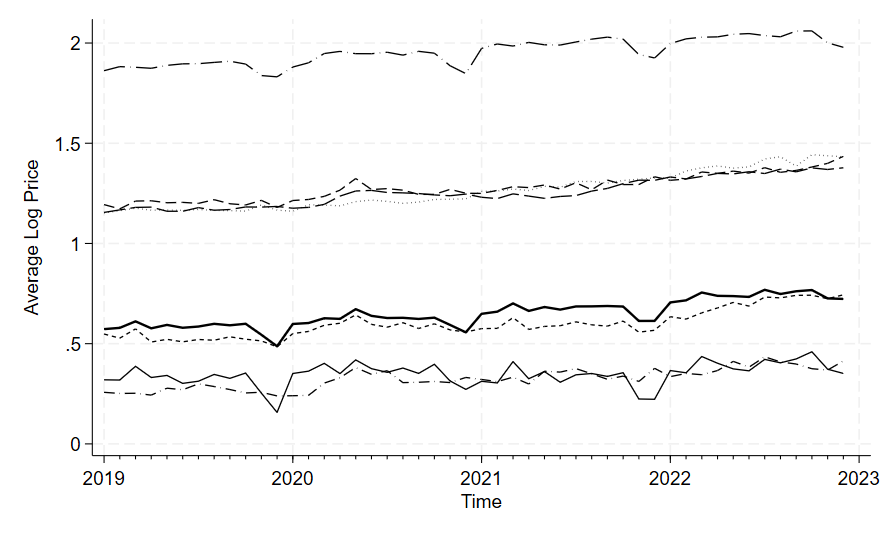}
    \end{minipage}
    \hfill
    \begin{minipage}[b]{0.49\textwidth}
        \centering
        \subcaption*{Panel B: Average share by department}
        \includegraphics[width=\linewidth]{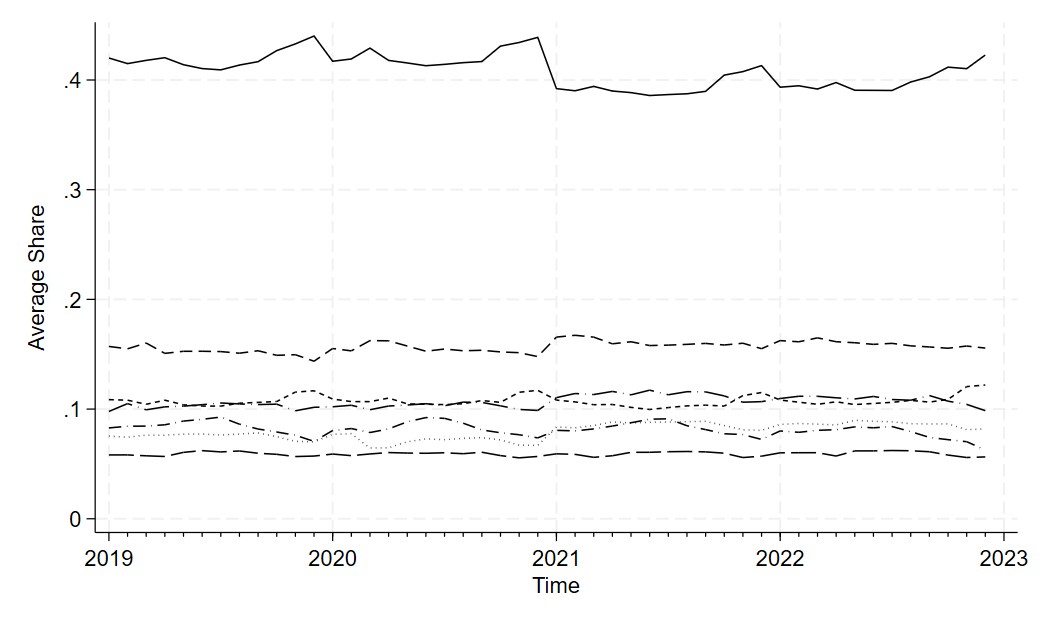}
    \end{minipage}
    \vspace{1em}
    \includegraphics[width=0.9\textwidth]{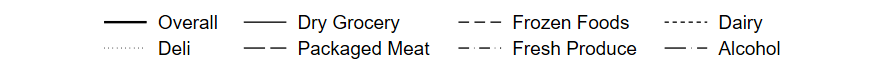}
            \caption{Time evolution of average log prices and average shares by department}
    \label{fig:logprice_share}

    \caption*{
       \makebox[\textwidth]{%
         \begin{minipage}{\textwidth}
               \justifying
               \noindent
               \footnotesize \textit{Notes:} Panel A reports the average log price of each department across time. The reported log price of each department corresponds to the average of household-specific Stone--Lewbel log price indices for that department in a given month. We compute the overall log price by weighting each household’s Stone--Lewbel log price indices by their own budget shares in a given month and averaging across households. Panel B reports the average share on each department across time.
        \end{minipage}
    }
    }
\end{figure}

Next, Figure \ref{fig:logprice_share} displays the evolution of average log prices and average shares across goods over time. Panel A shows the evolution of average log prices across goods over time, with temporary dips occurring around December. These seasonal declines are well documented and driven primarily by holiday promotions and year-end inventory clearance. Overall, prices rose rapidly after the onset of the COVID-$19$ pandemic in early $2020$, with cumulative inflation reaching approximately $10$\% by the end of $2020$ and an additional $10$\% by the end of $2022$. 

Panel B displays the evolution of average shares across goods over time. With the exception of dry grocery, which represents approximately $40\%$ of household food expenditures, the average share is around $10\%$ for any other good. Furthermore, though the shares appear relatively stable, we observe noticeable shifts during the COVID-$19$ period, especially for dry grocery and alcohol. While the other departments have more modest fluctuations, they still exhibit some adjustments in household spending. Taken together, the two panels show that household consumption adapted in response to the price changes during the pandemic, an important feature to consider when we turn to interpreting welfare impacts.

\subsection{Estimation and inference}\label{sec:estimation}

To obtain estimates for average welfare and its spread across the population, we need to model the first three moments of demand.\footnote{We do this using the consumer panel in pooled form, abstracting from its panel structure and treating all observations as a single cross section.} We recover those moments semiparametrically using a generalized additive model (GAM) in prices and expenditure.\footnote{A GAM generalizes generalized linear models by allowing for smooth functions of predictors. See \cite{Hastie1986} for an introduction to the theory and \cite{Wood2017} for an extensive review and details about its implementation in $R$.} Such an approach is flexible in budget sets while avoiding the curse of dimensionality. In our application, the GAM expresses moments of composite shares as the sum of smooth functions in prices and equivalized expenditure:
\[
W_{n}(\bp, y; \widetilde{\Delta \bp}) = \mathbb{E}[w_q^\omega(\bp, y; \Delta \bp)^n \mid \bp, y , \bx] =\sum_{j=1}^J f_{jn}(p_{j}) + g_{n}(y) + \bx^{\intercal} \boldsymbol{\beta}_{n},
\]
where $f_{jn}(\cdot)$ and $g_{n}(\cdot)$ are unknown smooth functions, $\bx$ is a set of control variables, and $\boldsymbol{\beta}_{n}$ is a vector of parameters.\footnote{We include age, race, and education as covariates to control for observable heterogeneity.} The functions $f_{jn}(\cdot)$ and $g_{n}(\cdot)$ are represented by means of basis expansions (e.g., cubic splines), and smoothness is controlled through penalization. The estimation then proceeds by minimizing a penalized likelihood:
\begin{align} \label{eq:gam}
\widehat{W_n}(\bp, y; \Delta \bp) 
&= \min_{f_{1n}, \dots, f_{Jn},\, g_n} \sum_{i=1}^n \left( 
    w_{iq}^\omega(\bp, y; \Delta \bp)^n 
    - \sum_{j=1}^J f_{jn}(x_{ij}) 
    - g_{n}(y_i) 
    - \bx_{i}^{\intercal} \boldsymbol{\beta}_{n} 
\right)^2 \notag \\
&\quad + \sum_{j=1}^J \lambda_{jn} \int \left( f_{jn}''(p) \right)^2 dp 
+ \lambda_{yn} \int \left( g_n''(y) \right)^2 dy,
\end{align}

\noindent
where the penalty terms $\lambda_n := (\lambda_{1n}, \lambda_{2n}, \dots, \lambda_{Jn}, \lambda_{yn})^{\intercal}$ control the trade-off between fit and smoothness and $h''(\cdot)$ denotes the second-order derivative of a function $h(\cdot)$. To penalize complexity more effectively, we use the restricted maximum likelihood (REML) approach to estimate the smoothing parameters $\lambda_{n}$.

Following the literature, we use equivalized household income $z$ as an instrument in a control function approach \citep{blundell_powell_2003, blundellmatzkin14}. Specifically, we regress total equivalized household expenditure on equivalized household income and prices:
\begin{equation*}
    y = \delta_0 + \delta_1 z + \mathbf{p}^\intercal \boldsymbol{\delta}_2 + \mathbf{x}^\intercal \boldsymbol{\delta}_3 + e.
\end{equation*}

\noindent
Then, we add the residual $\widehat{e}$ of this regression as an additional explanatory variable when estimating moments of the composite share. This approach helps eliminate the bias induced by endogeneity in our nonlinear moment equations. We apply our results using consistent plug-in estimators, and all confidence intervals are reported at the $95\%$ level and constructed from a nonparametric bootstrap with $199$ replications.

\subsection{Welfare analysis}

In what follows, we estimate the welfare impacts from the initial COVID-$19$ price shock, defined as the change in log prices faced by households between December $2020$ and December $2021$. Comparing log prices from the same month one year apart allows us to control for natural cyclical price variations and thus to isolate the impacts of the COVID-$19$ pandemic on prices. To allow meaningful interhousehold welfare comparisons, we fix the vector of baseline prices $(\bp_{0})$ to the average price in December $2020$ for every household. Finally, we set the change in log prices $(\widetilde{\Delta \bp})$ to the difference in average log prices between December $2020$ and December $2021$.\footnote{For Table \ref{tab:welfare_decomp_good}, we consider a single price change at a time such that the results for a good $j$ are obtained with $\widetilde{\Delta \bp} = (0, \dots, 0, \widetilde{\Delta p_j}, 0, \dots, 0)$. For Tables \ref{tab:welfare_decomp_composite}--\ref{tab:Bwelfare_decomp_composite}, we consider a joint price change such that $\widetilde{\Delta \bp} = (\widetilde{\Delta p_1}, \widetilde{\Delta p_2}, \dots, \widetilde{\Delta p_J})$.} When applicable, the price change is demographic specific to capture heterogeneity in prices faced arising from differences in location, store availability, and consumption behavior across demographic groups.


\paragraph{Average welfare effects.}

We begin our analysis by using Theorem \ref{thm:linearallmoments} to compute the CLI for each category of goods separately and to assess the bias that would be introduced from using a first-order approximation.\footnote{The first-order approximation is given by $\mathfrak{M}_1^{CLI}$ and represents the welfare impact of the price shock with the share held fixed at its baseline value.} This approach allows us to isolate the welfare impact of each individual price change, thereby clarifying the relative importance of different goods in driving the overall cost of living.
The results are reported in Table~\ref{tab:welfare_decomp_good}.
\begin{table}[H]
\centering
\caption{Average of CLI by department (\%)}
\label{tab:welfare_decomp_good}
\begin{adjustbox}{width=0.8\textwidth, center}
\begin{threeparttable}[b]
\begin{tabular}{l *{2}{C{2.75cm}} *{2}{C{2.75cm}}}
\toprule
Department & \multicolumn{2}{c}{Average of CLI} & \multicolumn{2}{c}{First-order approximation bias$^{a}$} \\
\cmidrule(lr){2-3} \cmidrule(lr){4-5}
      & Estimate & 95\% CI & Estimate & 95\% CI \\
\midrule
Dry Grocery     & 45.44  & [45.30, 45.58] & 10.82  & [10.55, 11.13] \\
\addlinespace
Frozen Foods    & 15.81  & [15.75, 15.86] &  1.01  & [0.73, 1.30] \\
\addlinespace
Dairy           &  10.93  & [10.90, 10.96] & 1.76  & [1.55, 1.95] \\
\addlinespace
Deli            & 8.33  & [8.27, 8.37] &  5.75  & [5.21, 6.25] \\
\addlinespace
Packaged Meat   &  6.03  & [6.01, 6.04] &  1.90  & [1.69, 2.11] \\
\addlinespace
Fresh Produce   & 8.33  & [8.31, 8.37] &  1.76  & [1.53, 1.99] \\
\addlinespace
Alcohol         & 15.29  & [15.20, 15.38] & 43.77  & [42.97, 44.55] \\
\bottomrule
\end{tabular}
\begin{tablenotes}
    \item[a] {\footnotesize The first-order approximation bias is given by 
    $\left((\mathfrak{M}_1^{CLI} + \mathfrak{B}_1^{CLI}) - \mathfrak{M}_1^{CLI} \right)/\mathfrak{M}_1^{CLI}$ evaluated at $(\bp_0, y; \widetilde{\Delta \bp})$.}
\end{tablenotes}
\end{threeparttable}
\end{adjustbox}
\end{table}

The first column of Table \ref{tab:welfare_decomp_good} shows significant heterogeneity in the average compensation needed for households to retain their pre-pandemic utility levels. On the higher end, a typical household would have had to be compensated by as much as $45.44\%$ of its total food expenditures given the price increase in dry grocery to keep its pre-pandemic utility constant. On the lower end, the required compensation is as little as $6.03\%$ for packaged meat. For other goods, the average compensation hovers around $10\%$ of households' total food expenditures. The relatively large CLI for dry grocery likely reflects that it accounts for a disproportionately large share of households' food expenditures, thus amplifying the welfare loss from its price increases.


Next, the second column of Table \ref{tab:welfare_decomp_good} presents the bias that would be introduced from using a first-order approximation of the CLI. Given the definition of our first-order approximation, a positive number means that the first-order approximation underestimates the true welfare loss, while a negative number means it overestimates the welfare loss. The results show that a first-order approximations can be highly misleading in estimating the welfare impacts across different categories of goods. In some cases, the first-order approximation dramatically underestimates the true welfare loss---for example, by approximately $44\%$ and $11\%$ for alcohol and dry grocery, respectively. These results highlight that the accuracy of the first-order approximation varies widely across goods, depending on income and substitution effects.

\paragraph{Welfare effects across observed heterogeneity.}

To better assess the welfare loss from the COVID-$19$ price shock, we next evaluate the changes in CLI using a composite share weighted by the change in the goods' log prices. The CLI for this composite share can be interpreted as a measure of a household's overall exposure to the price shock. Since a household's ability to mitigate price increases through substitution or price search may differ systematically across demographics, we consider demographic-specific price shocks that reflect the change in average log price faced by each demographic group.\footnote{The price shock used for the row ``All" in Tables \ref{tab:welfare_decomp_composite}--\ref{tab:Bwelfare_decomp_composite} is the average change in log price in the whole sample.} Similarly to Table \ref{tab:welfare_decomp_good} above, Table \ref{tab:welfare_decomp_composite} presents the results for the CLI and the bias from the first-order approximation.
\begin{table}[H]
\captionof{table}{Average of CLI by race and education ($\%$)} 
\label{tab:welfare_decomp_composite}
\begin{adjustbox}{max width=0.9\textwidth, center}
\begin{threeparttable}[b]
  \begin{tabular}{ll *{2}{C{2.75cm}} *{2}{C{2.75cm}}}
    \toprule
    \multirow{2}{*}{Race$^a$} & \multirow{2}{*}{Education level$^b$}
    & \multicolumn{2}{c}{Average of CLI} 
    & \multicolumn{2}{c}{First-order approximation bias$^{c}$} \\
    \cmidrule(lr){3-4} \cmidrule(lr){5-6}
    & & Estimate & 95\% CI & Estimate & 95\% CI \\
    \midrule

    \multirow{3}{*}{White}
    & Low   & 8.21 & [8.20, 8.21] & 4.23  & [4.19, 4.28] \\
    & Mid   & 9.56 & [9.55, 9.56] & 4.72  & [4.67, 4.77] \\
    & High  & 9.62 & [9.62, 9.63] & 4.34 & [4.28, 4.40] \\
\addlinespace
    \multirow{3}{*}{Black}
    & Low   & 5.44 & [5.43, 5.45] & 3.16 & [3.01, 3.29] \\
    & Mid   & 4.71 & [4.70, 4.72] & 1.60 & [1.46, 1.73] \\
    & High  & 9.79 & [8.78, 8.81] & 4.32 & [4.20, 4.43] \\
\addlinespace
    \multirow{3}{*}{Asian}
    & Low   & 13.86 & [13.85, 13.87] & 5.45 & [5.38, 5.52] \\
    & Mid   & 4.50 & [4.48, 4.51] & -1.67 & [-1.88, -1.43] \\
    & High  & 11.82 & [11.80, 11.85] & 3.38 & [3.22, 3.54] \\
\addlinespace
    \multirow{3}{*}{Other}
    & Low   & -1.59 & [-1.60, -1.57] & 5.32 & [4.70, 5.93] \\
    & Mid   & 8.70 & [8.69, 8.70] & 4.31 & [4.26, 4.36] \\
    & High  & 2.85 & [2.84, 2.86] & -1.41 & [-1.65, -1.12] \\
\addlinespace
    \midrule
    \multicolumn{2}{l}{All} & 8.47 & [8.46, 8.48] & 4.28 & [4.23, 4.34] \\
    \bottomrule
  \end{tabular}

  \begin{tablenotes}
    \item[] {\footnotesize \textit{Notes:} Welfare effects are computed from conditional moment estimates evaluated at the sample median age while allowing other covariates $\bx$ to vary.}
    \item[a] {\footnotesize Race represents the racial identity of the household as reported by the panelist.}
    \item[b] {\footnotesize Education level refers to the highest level of education attained by any household head. Low = high school diploma or less; Mid = some college or college degree; High = postgraduate degree.}
    \item[c] {\footnotesize The first-order approximation bias is given by $\left((\mathfrak{M}_1^{CLI} + \mathfrak{B}_1^{CLI}) - \mathfrak{M}_1^{CLI} \right)/\mathfrak{M}_1^{CLI}$ evaluated at $(\bp_0, y; \widetilde{\Delta \bp})$.}
  \end{tablenotes}
\end{threeparttable}
\end{adjustbox}
\end{table}

The first column shows that the average CLI is $8.47\%$ across the full sample. This implies that, on average, a household would have required approximately $108\%$ of its December $2020$ food expenditures in December $2021$ to maintain the same utility level. Notably, Table \ref{tab:welfare_decomp_composite} also reveals considerable heterogeneity in CLI values, with point estimates ranging from $-1.59\%$ for Other--low education households to $13.86\%$ for Asian--low education households. The negative CLI for the former subgroup is driven by a sharp decline in the average price of dry grocery\textemdash the department with the largest budget share. Overall, there is no clear pattern in the CLIs across race and education, suggesting that the welfare impacts of the COVID-19 price shock were not systematically linked to these observable characteristics.

The second column shows that the bias from the first-order approximation is $4.28\%$ on average, meaning that the approximation tends to underestimate the welfare losses by approximately $4\%$. As with the CLI, there is substantial heterogeneity in the magnitude of this bias. For example, Asian households with low and mid education levels have a bias of approximately $5\%$ and $-2\%$, respectively, while the bias for White households is at around $4\%$. These disparities reflect significant price and income effects, which cause the true welfare response to diverge from the first-order approximation.

\paragraph{Welfare effects across unobserved heterogeneity.}
This section looks into variation in the cost of living within race--education subgroups to uncover heterogeneity driven by unobservable characteristics. To this end, we compute the standard deviation of the CLI from the first two moments of the CLI provided by Theorem \ref{thm:linearallmoments}. The results for each subgroup are reported in Table \ref{tab:variance_cli}.
\begin{table}[H]
\captionof{table}{Standard deviation of CLI by race and education} 
\label{tab:variance_cli}
\begin{adjustbox}{max width=0.62\textwidth, center}
\begin{threeparttable}[b]
\begin{tabular}{ll *{2}{C{3cm}}}
\toprule
\multirow{2}{*}{Race$^{a}$} & \multirow{2}{*}{Education level$^{b}$} & \multicolumn{2}{c}{Standard deviation of CLI} \\
\cmidrule(lr){3-4}
 &  & Estimate & 95\% CI \\
\midrule
\multirow{3}{*}{White} 
 & Low  & 1.00 & [1.00, 1.01] \\
 & Mid  & 1.03 & [1.03, 1.04] \\
 & High & 1.37 & [1.37, 1.38] \\
\addlinespace
\multirow{3}{*}{Black} 
 & Low  & 1.91 & [1.90, 1.92] \\
 & Mid  & 1.99 & [1.98, 2.00] \\
 & High & 2.21 & [2.20, 2.22] \\
\addlinespace
\multirow{3}{*}{Asian} 
 & Low  & 1.08 & [1.07, 1.10] \\
 & Mid  & 2.40 & [2.39, 2.41] \\
 & High & 2.68 & [2.66, 2.69] \\
\addlinespace
\multirow{3}{*}{Other} 
 & Low  & 2.89 & [2.88, 2.90] \\
 & Mid  & 1.00 & [1.00, 1.01] \\
 & High & 2.01 & [2.01, 2.02] \\
\midrule
All & & 1.17 & [1.16, 1.17] \\
\bottomrule
\end{tabular}

\begin{tablenotes}
  \item[] {\footnotesize \textit{Notes:} We compute the standard deviations from conditional moment estimates evaluated at the sample median age while allowing other covariates $\bx$ to vary.} \setstretch{1.0}
  \item[a] {\footnotesize Race represents the racial identity of the household as reported by the panelist.} \setstretch{1.0}
  \item[b] {\footnotesize Education level refers to the highest level of education attained by any household head. Low = high school diploma or less; Mid = some college or college degree; High = postgraduate degree.} \setstretch{1.0}
\end{tablenotes}

\end{threeparttable}
\end{adjustbox}
\end{table}

Table \ref{tab:variance_cli} reveals notable within-group variation in the CLIs across race--education subgroups. The standard deviations range from $1.00$ to $2.89$, generally with greater dispersion for higher education levels. For example, the standard deviations of the CLI are $2.21$ and $2.68$ for Black and Asian households with high education levels while only $1.91$ and $1.08$ for Black and Asian households with low education levels. These results suggest that households within the same subgroups experienced markedly different welfare impacts from the price changes, particularly among the more highly educated. 

\begin{figure}[H]
    \centering
    \includegraphics[width=0.68\textwidth]{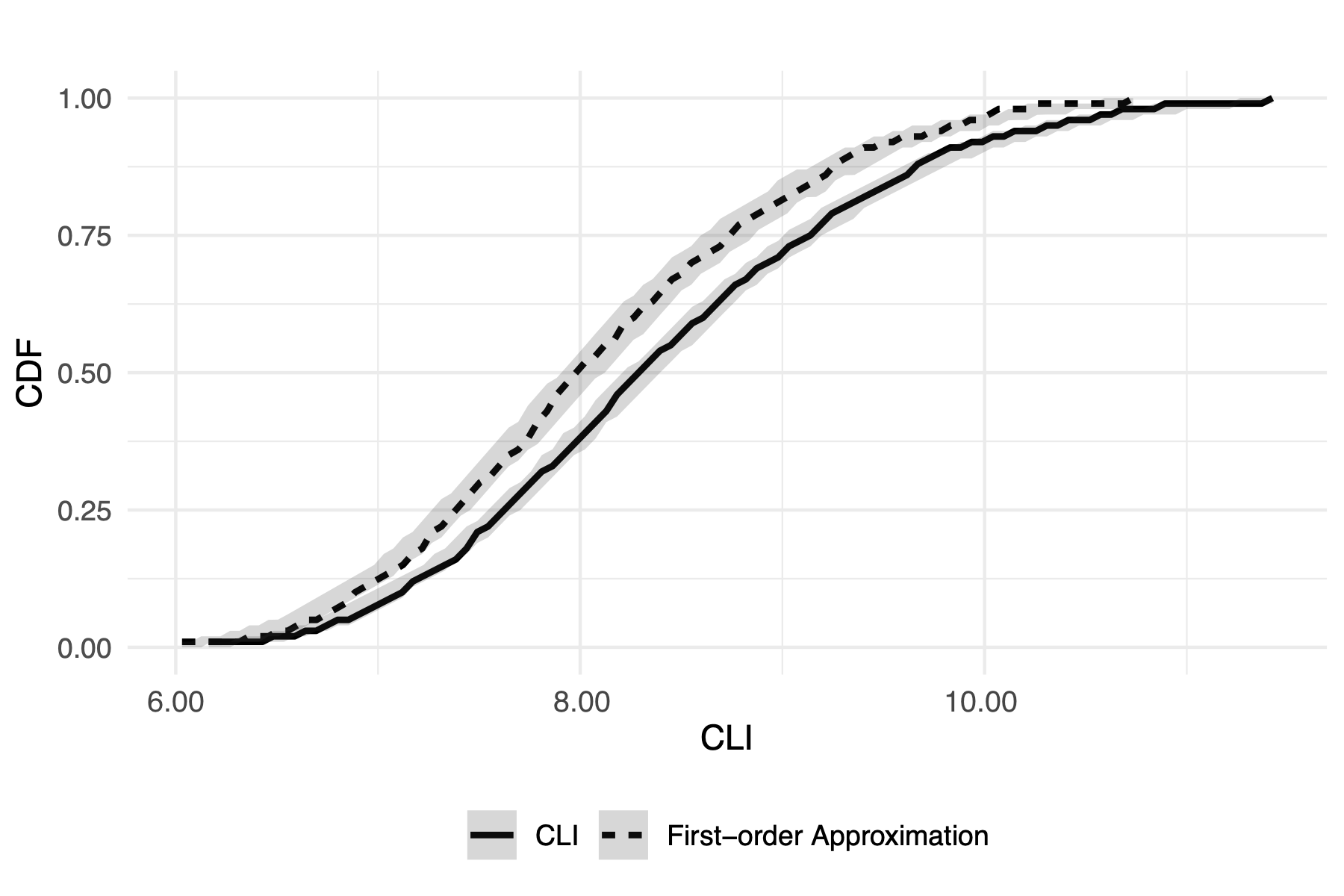}
        \caption{Cumulative distributions of CLI and first-order approximation}
    \label{fig:cdf}
    \caption*{
        \makebox[\textwidth]{%
            \begin{minipage}{0.68\textwidth}
                \justifying
                \noindent
                \footnotesize \textit{Notes:} The cumulative distributions of the CLI and of the first-order approximation are obtained from a standard local linear conditional quantile estimation that includes prices, total equivalized expenditure, and the error from the control function approach as regressors. The price level is set to the average price in December 2020, the total equivalized expenditure level to the sample median, and the error to zero.
            \end{minipage}
        }
    }
\end{figure}

To gain further insight into the heterogeneity in the cost of living, we exploit Proposition \ref{prop:quantile} to recover the CDFs of both the CLI and the first-order approximation. The CDFs are displayed in Figure \ref{fig:cdf}, where the horizontal axis shows the required percentage increase in total expenditure needed to achieve the utility level from December $2020$ in the face of the joint price shock from December $2020$ to December $2021$. The vertical axis shows the share of households whose welfare loss is at or below each value.

Consistent with our previous results, the two curves in Figure \ref{fig:cdf} reveal that the first-order approximation systematically underestimates the welfare loss and that most households are within $2$ percentage point of the average CLI.

\paragraph{Sensitivity to preference restrictions.}

In this section, we take advantage of Proposition \ref{cor:decomposition} to analyze the sensitivity of the CLI across demographics to various restrictions on preferences. Table \ref{tab:Bwelfare_decomp_composite} reports the decomposition of the behavioral component of the CLI across race--education subgroups.

\begin{table}[H]
\captionof{table}{Behavioral decomposition of CLI by race and education $(\%)$} 
\label{tab:Bwelfare_decomp_composite}
\begin{adjustbox}{max width=\textwidth, center}
\begin{threeparttable}[b]
  \begin{tabular}{ll *{6}{C{2.7cm}}}
    \toprule
    \multirow{2}{*}{Race$^a$} & \multirow{2}{*}{Education level$^b$}
    & \multicolumn{2}{c}{$\mathfrak{D}_{1,1} / \mathfrak{B}_1^{CLI}$} 
    & \multicolumn{2}{c}{$(\mathfrak{D}_{1,1} + \mathfrak{D}_{1,2}) / \mathfrak{B}_1^{CLI}$} 
    & \multicolumn{2}{c}{$(\mathfrak{D}_{1,1} + \mathfrak{D}_{1,3}) / \mathfrak{B}_1^{CLI}$} \\
    \cmidrule(lr){3-4} \cmidrule(lr){5-6} \cmidrule(lr){7-8}
    & & Estimate & 95\% CI & Estimate & 95\% CI & Estimate & 95\% CI \\
    \midrule

    \multirow{3}{*}{White}
    & Low   & 97.16 & [96.48, 97.88] & 99.05 & [98.99, 99.12] & 98.21 & [97.52, 98.93] \\
    & Mid   & 98.52 & [97.84, 99.21] & 99.17 & [99.11, 99.23] & 99.41 & [98.72, 100.10] \\
    & High  & 97.72 & [96.84, 98.64] & 98.47 & [98.36, 98.57] & 99.25 & [98.36, 100.18] \\
\addlinespace
    \multirow{3}{*}{Black}
    & Low   & 92.71 & [91.15, 94.52] & 93.29 & [92.67, 93.81] & 99.92 & [98.20, 101.87] \\
    & Mid   & 73.92 & [69.73, 77.43] & 80.12 & [77.73, 81.87] & 92.61 & [88.10, 96.58] \\
    & High  & 92.17 & [90.56, 93.64] & 95.59 & [95.29, 95.88] & 96.55 & [94.94, 98.16] \\
\addlinespace
    \multirow{3}{*}{Asian}
    & Low   & 99.66 & [99.09, 100.20] & 99.31 & [99.25, 99.36] & 100.29 & [99.72, 100.84] \\
    & Mid   & 118.85 & [113.56, 124.15] & 121.78 & [119.01, 126.12] & 95.64 & [91.56, 99.36] \\
    & High  & 95.78 & [94.02, 97.90] & 95.41 & [94.98, 95.81] & 100.60 & [98.80, 102.90] \\
\addlinespace
    \multirow{3}{*}{Other}
    & Low   & 134.73 & [130.03, 141.28] & 138.29 & [132.99, 144.73] & 95.91 & [93.54, 98.63] \\
    & Mid   & 96.89 & [96.19, 97.60] & 99.06 & [90.00, 99.12] & 97.85 & [97.15, 98.58] \\
    & High  & 138.14 & [129.89, 149.32] & 134.66 & [129.04, 143.86] & 103.84 & [98.81, 109.67] \\

    \midrule
    \multicolumn{2}{l}{All} & 97.37 & [96.53, 98.22] & 98.72 & [98.63, 98.81] & 98.71 & [97.86, 99.58] \\
    \bottomrule
  \end{tabular}
  
  \begin{tablenotes}
    \item[]{\footnotesize \textit{Notes:} The first column reports the estimate for the behavioral component of the CLI when preferences are assumed to be homogeneous and homothetic ($\mathfrak{D}_{1,1}$), the second the estimate when preferences are homogeneous and nonhomothetic ($\mathfrak{D}_{1,1} + \mathfrak{D}_{1,2}$), and the last the estimate when preferences are heterogeneous and homothetic ($\mathfrak{D}_{1,1} + \mathfrak{D}_{1,3}$), each normalized relative to the total behavioral response $\mathfrak{B}_1^{CLI}$. We compute welfare effects from conditional moment estimates evaluated at the sample median age while allowing other covariates $\bx$ to vary.}
    \setstretch{1.0}
    \item[a] {\footnotesize Race represents the racial identity of the household as reported by the panelist.} \setstretch{1.0}
    \item[b] {\footnotesize Education level refers to the highest level of education attained by any head household member. Low = high school diploma or less; Mid = some college or college degree; High = postgraduate degree.} \setstretch{1.0}
  \end{tablenotes}

\end{threeparttable}
\end{adjustbox}
\end{table}

Table \ref{tab:Bwelfare_decomp_composite} shows that every model is within $3\%$ of the true behavioral response when applied to the whole sample ("All"). At the subgroup level, however, the specification with homogeneous and homothetic preferences $(\mathfrak{D}_{1,1})$ does not perform as well. The estimated response is too low at $73.92\%$ for Black households with mid education, and too high at $138.14\%$ for Other households with high education. Allowing for nonhomotheticity while maintaining homogeneity $(\mathfrak{D}_{1,1} + \mathfrak{D}_{1,2})$ improves the estimate for Black households with mid education to $80.12\%$, but in other cases it further inflates the response, reaching $138.29\%$ for Other households with low education. By contrast, introducing heterogeneity while maintaining homotheticity $(\mathfrak{D}_{1,1} + \mathfrak{D}_{1,3})$ yields consistent improvements across all race–education subgroups, with behavioral responses deviating by no more than $5\%$ from the full behavioral response. Taken together, these results highlight heterogeneity as the key first-order feature required to accurately capture cost-of-living adjustments.

\section{Conclusion}\label{sec:conclusion}

In this paper, we introduce a novel method to identify the curvature of the expenditure function through moments of demand. Leveraging this insight can enable more accurate counterfactual exercises in applied welfare analysis. Importantly, we establish that no better approximations can be derived from cross-sectional data. Looking ahead, we highlight several promising directions for future research. First, it would be valuable to investigate the  additional identifying power short consumer panels provide for estimating counterfactuals. Initial steps in this direction have been taken by \citet*{crawford19}, \citet*{coopriderhoderleinmeister}, and \citet*{chernozhukovetal}. Second, extending our approach to general equilibrium models could improve the measurement of welfare gains and losses from trade, as explored in \citet{baqaeeburstein}. Third, further work is needed to deepen our understanding of welfare analysis in models that depart from standard rationality assumptions. Recent contributions by \citet*{apesteguiaballester} and \citet*{aguiarserrano} provide a foundation for such research. Finally, efficient estimation in high-dimensional settings with many goods remains an open challenge. Advances in high-dimensional statistics offer potential solutions to the curse of dimensionality, a problem further exacerbated by the endogeneity of budget sets. In this regard, building on the results of \citet*{chernozhukovhausmannewey} appears to be a particularly promising avenue for future study.

\newpage
\renewcommand{\bibfont}{\small}
\setlength{\bibsep}{2pt}
\bibliography{Raghav}

\clearpage
\setcounter{page}{1}
\renewcommand{\thepage}{A\arabic{page}}

\begin{center}
    {\huge Appendix \\
    \vspace{10pt}
    \Large \textit{Consumer Welfare Under Individual Heterogeneity}
  }
\end{center}

\appendix
\small 

\section{Proofs omitted in the main text}\label{app:proofs}

\subsection{Proof of Proposition~\ref{thm:linearcompe} for budget shares}
We first provide two intermediate results; Lemma~\ref{lem:slutskyshares} reformulates the Slutsky equation in terms of budget shares and Lemma~\ref{lem:incomeshares} adapts Lemma~\ref{lem:income} to moments of budget shares.
\begin{lem}\label{lem:slutskyshares}
    For all $(\bp, y) \in \mathcal{P}\times \mathcal{Y}$, it holds that
    \begin{equation*}
        D_{\widetilde{\bp} }\mathbf{w}_h^\omega(\bp, v^\omega(\bp, y)) = D_{\widetilde{\bp} }\mathbf{w}_q^\omega(\bp, y) + D_{\widetilde{y}}\mathbf{w}_q^\omega(\bp, y) \mathbf{w}_q^\omega(\bp, y)^\intercal + \mathbf{w}_q^\omega(\bp, y) \mathbf{w}_q^\omega(\bp, y)^\intercal.
    \end{equation*}
\end{lem}
\begin{proof}
    Observe that
    \begin{equation*}
        \begin{split}
            D_{\widetilde{\bp} }\mathbf{w}_q^\omega(\bp, y) &= D_{\widetilde{\bp} } \frac{\bp \odot \bq^\omega(\bp, y)}{y} \\
            &= \frac{1}{y}\bp \bp^\intercal \odot D_\bp \bq^\omega(\bp,y) + \text{diag}(\mathbf{w}_q^\omega(\bp, y)), \\
        \end{split}
    \end{equation*}
    and 
    \begin{equation*}
        \begin{split}
            D_{\widetilde{y}}\mathbf{w}_q^\omega(\bp, y) \mathbf{w}_q^\omega(\bp, y)^\intercal &= D_{\widetilde{y} } \frac{\bp \odot \bq^\omega(\bp, y)}{y} \mathbf{w}_q^\omega(\bp, y)^\intercal \\
            &= [\bp \odot D_y \bq^\omega(\bp, y) - \mathbf{w}_q^\omega(\bp, y)] \mathbf{w}_q^\omega(\bp, y)^\intercal \\
            &= \frac{1}{y}\bp \bp^\intercal \odot D_y \bq^\omega(\bp,y) \bq^\omega(\bp,y) ^\intercal - \mathbf{w}_q^\omega(\bp, y) \mathbf{w}_q^\omega(\bp, y)^\intercal.
        \end{split}
    \end{equation*}
    Therefore, it holds that
    \begin{equation*}
        \begin{split}
            D_{\widetilde{\bp} }\mathbf{w}_h^\omega(\bp, v^\omega(\bp, y)) &= D_{\widetilde{\bp} } \frac{\bp \odot \bh^\omega(\bp, v^\omega(\bp, y))}{y} \\
            &= \frac{1}{y}\bp \bp^\intercal \odot D_\bp \bh^\omega(\bp,v^\omega(\bp, y)) + \text{diag}(\mathbf{w}_h^\omega(\bp, v^\omega(\bp, y))) \\
            &= \frac{1}{y}\bp \bp^\intercal \odot \left[D_\bp \bq^\omega(\bp,y) + D_y \bq^\omega(\bp,y) \bq^\omega(\bp,y) ^\intercal  \right]+ \text{diag}(\mathbf{w}^\omega_q(\bp, y)) \\
            &= D_{\widetilde{\bp} }\mathbf{w}_q^\omega(\bp, y) + D_{\widetilde{y}}\mathbf{w}_q^\omega(\bp, y) \mathbf{w}_q^\omega(\bp, y)^\intercal + \mathbf{w}_q^\omega(\bp, y) \mathbf{w}_q^\omega(\bp, y)^\intercal,
        \end{split}
    \end{equation*}
    where the third equality follows from the Slutsky equation.
\end{proof}

\begin{lem}\label{lem:incomeshares}
     For every $n \in \mathbb{N}_{++}$, $\bt \in \mathbb{R}^k$ and $(\bp, y) \in \mathcal{P}\times \mathcal{Y}$, it holds that 
    \begin{equation*}
        \begin{split}
            \mathbb{E}\left[w_q^\omega(\bp,y; \bt)^{n-1}  D_{\widetilde{\bp}} w_q^\omega(\bp,y; \bt)\right] &= \frac{1}{n}  D_{\widetilde{\bp}} W_{n}(\bp,y; \bt), \\
            \mathbb{E}\left[w_q^\omega(\bp,y; \bt)^{n-1}  D_{\widetilde{y}} w_q^\omega(\bp,y; \bt)\right] &= \frac{1}{n}  D_{\widetilde{y}} W_{n}(\bp,y; \bt).
        \end{split}
    \end{equation*}
\end{lem}
\begin{proof}
    The proof is omitted as it is analogous to that of Lemma~\ref{lem:income}, mutatis mutandis.
\end{proof}

We are now equipped to prove Proposition~\ref{thm:linearcompe} for budget shares. Using Lemma~\ref{lem:slutskyshares}, for any $(\bp, y) \in \mathcal{P}\times \mathcal{Y}$, we have
    \begin{equation*}
        \begin{split}
            \widetilde{c}^\omega(\bp, y; \bt) &
        = \bt^\intercal D_{\widetilde{\bp}}^2  \widetilde{e}^\omega(\bp, v^\omega(\bp, y)) \bt \\
            &= \bt^\intercal D_{\widetilde{\bp}} \mathbf{w}_h^\omega(\bp, v^\omega(\bp, y)) \bt \\
            &= \bt^\intercal \left[D_{\widetilde{\bp} }\mathbf{w}_q^\omega(\bp, y) + D_{\widetilde{y}}\mathbf{w}_q^\omega(\bp, y) \mathbf{w}_q^\omega(\bp, y)^\intercal + \mathbf{w}_q^\omega(\bp, y) \mathbf{w}_q^\omega(\bp, y)^\intercal\right]\bt \\
            &= D_{\widetilde{\bp}}(\bt^\intercal \mathbf{w}_q^\omega(\bp, y))\bt + (\bt^\intercal \mathbf{w}_q^\omega(\bp, y))D_{\widetilde{y}}(\bt^\intercal \mathbf{w}_q^\omega(\bp, y)) + (\bt^\intercal \mathbf{w}_q^\omega(\bp, y))^2 \\
            &= D_{\widetilde{\bp}}w_q^\omega(\bp,y; \bt)\bt + w_q^\omega(\bp,y; \bt)D_{\widetilde{y}}w_q^\omega(\bp,y; \bt) + w_q^\omega(\bp,y; \bt)^2, \\
        \end{split}
    \end{equation*}
    where the second equality follows from Shephard's lemma for budget shares. Therefore, 
    \begin{equation*}
        \begin{split}
            \mathbb{E}[w^\omega_q(\bp, y; \bt)^{n-1} \widetilde{c}^\omega(\bp, y; \bt)] & = \mathbb{E} [w^\omega_q(\bp, y; \bt)^{n-1}[ D_{\widetilde{\bp}}w_q^\omega(\bp,y; \bt)\bt + w_q^\omega(\bp,y; \bt)D_{\widetilde{y}}w_q^\omega(\bp,y; \bt) + w_q^\omega(\bp,y; \bt)^2]] \\
            & = \frac{1}{n} D_{\widetilde{\bp}} W_n(\bp, y; \bt)\bt + \frac{1}{n+1} D_{\widetilde{y}} W_{n+1}(\bp, y; \bt) + W_{n+1}(\bp, y; \bt),
        \end{split}
    \end{equation*}
     where the last equality follows from Lemma~\ref{lem:incomeshares}.

\subsection{Proof of Theorem~\ref{thm:linearallmoments} for the cost-of-living index}\label{app:prooflinearallmoments}
The proof for the moments of the CLI is similar to that of the moments of the CV. The second-order expansion of the log expenditure function for $\widetilde{\Delta \bp} \approx \mathbf{0}$ can be written as
\begin{equation*}
        \begin{split}
            \widetilde{e}^\omega({\bp}_1, v_0^\omega) &= \widetilde{e}^\omega({\bp}_0, v_0^\omega) + D_{\widetilde{\bp}} \widetilde{e}^\omega({\bp}_0, v_0^\omega) \widetilde{\Delta \bp} + \frac{1}{2}(\widetilde{\Delta \bp})^\intercal D_{\widetilde{\bp}}^2 \widetilde{e}^\omega({\bp}_0, v_0^\omega) \widetilde{\Delta \bp} + O(||\widetilde{\Delta \bp}||^3) \\
            &= \widetilde{y} + w^\omega_q(\bp_0, y; \widetilde{\Delta \bp}) + \frac{1}{2}\widetilde{c}^\omega(\bp_0, y; \widetilde{\Delta \bp}) + O(||\widetilde{\Delta \bp}||^3).
        \end{split}
    \end{equation*}
Therefore, using Equation~\eqref{eq:COLI}, the second-order approximation of the CLI becomes
    \begin{equation*}
        CLI^\omega(\bp_0, \bp_1, y) = w^\omega_q(\bp_0, y; \widetilde{\Delta \bp}) + \frac{1}{2}\widetilde{c}^\omega(\bp_0, y; \widetilde{\Delta \bp}) + O(||\widetilde{\Delta \bp}||^3),
    \end{equation*}
    such that for higher powers we obtain
    \begin{equation*}
        CLI^\omega(\bp_0, \bp_1, y)^n = w^\omega_q(\bp_0, y; \widetilde{\Delta \bp})^n + \frac{n}{2} w^\omega_q(\bp_0, y; \widetilde{\Delta \bp})^{n-1}\widetilde{c}^\omega(\bp_0, y; \widetilde{\Delta \bp}) + O(||\widetilde{\Delta \bp}||^{n+2}).
    \end{equation*}
    Taking expectations on both sides and using Proposition~\ref{thm:linearcompe} gives
    \begin{equation*}
        \begin{split}
            \mathbb{E}[CLI^\omega(\bp_0, \bp_1, y)^n] =\mathfrak{M}_n^{CLI}(\bp_0, y; \widetilde{\Delta \bp}) +  \mathfrak{B}_n^{CLI}(\bp_0, y; \widetilde{\Delta \bp}) + O(||\widetilde{\Delta \bp}||^{n+2}),
        \end{split}
    \end{equation*}
    as desired.

\subsection{Proof of Proposition~\ref{prop:quantile}}

\begin{lem}\label{lem:hoderleinlemma}
    Under standard regularity conditions \citep{hoderleinmammen}, we have that
\begin{equation*}
    \begin{split}
        K_{q, \tau}(\bp, y ; \bt)^{n-1} S_{\tau}^{CV}(\bp, y ; \bt) &= \frac{1}{2}\mathbb{E} \left[q^\omega(\bp, y; \bt)^{n-1} c^\omega(\bp, y; \bt) \mid  q^\omega(\bp, y; \bt) = K_{q, \tau}(\bp, y ; \bt)\right], \\
        K_{w, \tau}(\bp, y ; \bt)^{n-1} S_\tau^{CLI}(\bp, y; \bt) &= \frac{1}{2} \mathbb{E} \left[ w^\omega_q(\bp, y; \bt)^{n-1}\widetilde{c}^\omega(\bp, y; \bt) \mid w^\omega_q(\bp, y; \bt) = K_{w, \tau}(\bp, y; \bt)\right],
    \end{split}
\end{equation*}
    where
    \begin{equation*}
        \begin{split}
            S_\tau^{CV}(\bp, y; \bt) &= \frac{1}{2}  \left(D_\bp K_{q, \tau}(\bp, y ; \bt)  \bt +  K_{q, \tau}(\bp, y ; \bt) D_y K_{q, \tau}(\bp, y ; \bt) \right), \\
            S_\tau^{CLI}(\bp, y; \bt) &= \frac{1}{2} \left(D_{\widetilde{\bp}}K_{w, \tau}(\bp, y ; \bt)\bt + K_{w, \tau}(\bp, y ; \bt)D_{\widetilde{y}}K_{w, \tau}(\bp, y ; \bt) + K_{w, \tau}(\bp, y ; \bt)^2\right). \\
        \end{split}
    \end{equation*}
\end{lem}
\begin{proof}
    This result builds on the local average structural derivative characterization by \citet{hoderleinmammen}. We present the argument for the CV; the case of the CLI follows analogously. 

    From the proof of Proposition~\ref{thm:linearcompe}, we have
    \begin{equation*}
        \begin{split}
            c^\omega(\bp, y; \bt) &= D_\bp q^\omega(\bp, y; \bt)\bt + q^\omega(\bp, y; \bt)D_y q^\omega(\bp, y; \bt).
        \end{split}
    \end{equation*}
    Hence,
    \begin{equation*}
    \begin{split}
        &\mathbb{E} \left[q^\omega(\bp, y; \bt)^{n-1} c^\omega(\bp, y; \bt) \mid  q^\omega(\bp, y; \bt) = K_{q, \tau}(\bp, y ; \bt)\right] \\
        &\qquad = \mathbb{E} \left[q^\omega(\bp, y; \bt)^{n-1} D_\bp q^\omega(\bp, y; \bt)\bt \mid  q^\omega(\bp, y; \bt) = K_{q, \tau}(\bp, y ; \bt)\right] \\
        &\qquad \qquad + \mathbb{E} \left[q^\omega(\bp, y; \bt)^{n} D_y q^\omega(\bp, y; \bt)\mid  q^\omega(\bp, y; \bt) = K_{q, \tau}(\bp, y ; \bt)\right] \\
        &\qquad = K_{q, \tau}(\bp, y ; \bt)^{n-1}D_\bp K_{q, \tau}(\bp, y ; \bt)\bt + K_{q, \tau}(\bp, y ; \bt)^n D_y K_{q, \tau}(\bp, y ; \bt) \\
        &\qquad = 2 K_{q, \tau}(\bp, y ; \bt)^{n-1} S_{\tau}^{CV}(\bp, y; \bt),
    \end{split}
\end{equation*}
where the second equality follows from \citet{hoderleinmammen}.\footnote{\citet{Dettehoderlein2016} obtain a similar result in their Theorem 1.}
\end{proof}

\paragraph{Compensating variation.}
    For each quantile $\tau \in (0,1)$ of composite demand, define the population object
    \begin{equation*}
        \begin{split}
                R^{CV}_\tau(\bp_0, y; \Delta \bp) &:= K_{q, \tau}(\bp_0, y ; \Delta \bp) + S_{\tau}^{CV}(\bp_0, y; \Delta \bp). \\
        \end{split}
    \end{equation*}
     Intuitively, this object can be interpreted as the second-order approximation of the CV for a hypothetical consumer at the $\tau$th quantile of composite demand. Analogous to Theorem~\ref{thm:linearallmoments}, we obtain
    \begin{equation*}
        R^{CV}_\tau(\bp_0, y; \Delta \bp)^n = K_{q, \tau}(\bp_0, y ; \Delta \bp)^n + n K_{q, \tau}(\bp_0, y ; \Delta \bp)^{n-1} S_{\tau}^{CV}(\bp_0, y; \Delta \bp) + O(||\Delta \bp||^{n+2}).
    \end{equation*}
    Since $\tau$ is uniformly distributed, we have
    \begin{equation*}
        \begin{split}
            \mathbb{E}_\tau\left[R^{CV}_\tau(\bp_0, y; \Delta \bp)^n\right] &= \int_0^1 R^{CV}_\tau(\bp_0, y; \Delta \bp)^n d \tau \\
            &= \int_0^1 \left[K_{q, \tau}(\bp_0, y ; \Delta \bp)^n + n K_{q, \tau}(\bp_0, y ; \Delta \bp)^{n-1} S_{\tau}^{CV}(\bp_0, y; \Delta \bp)
            \right] d\tau + O(||\Delta \bp||^{n+2}) \\
            &= \mathfrak{M}_n^{CV}(\bp_0, y; \Delta \bp) + \mathfrak{B}_n^{CV}(\bp_0, y; \Delta \bp) +O(||\Delta \bp||^{n+2}),
        \end{split}
    \end{equation*}
    where the last equality follows from Lemma~\ref{lem:hoderleinlemma}. That is, the $n$th moment of $R^{CV}_\tau(\bp_0, y; \Delta \bp)$ coincides with that of the CV (as derived in Theorem~\ref{thm:linearallmoments}), up to order $n+1$.

    \paragraph{Cost-of-living index.}
    For each quantile $\tau \in (0,1)$ of composite budget share, define the population object
    \begin{equation*}
        R^{CLI}_\tau(\bp_0, y; \widetilde{\Delta \bp}) := K_{w, \tau}(\bp_0, y ; \widetilde{\Delta \bp}) + S_\tau^{CLI}(\bp_0, y; \widetilde{\Delta \bp}).
    \end{equation*}
    Analogous to Theorem~\ref{thm:linearallmoments}, we obtain
    \begin{equation*}
        R^{CLI}_\tau(\bp_0, y; \widetilde{\Delta \bp})^n = K_{w, \tau}(\bp_0, y ; \widetilde{\Delta \bp})^n + n K_{w, \tau}(\bp_0, y ; \widetilde{\Delta \bp})^{n-1} S_\tau^{CLI}(\bp_0, y; \widetilde{\Delta \bp}) + O(||\widetilde{\Delta \bp}||^{n+2}).
    \end{equation*}
    Since $\tau$ is uniformly distributed, we have
    \begin{equation*}
        \begin{split}
            \mathbb{E}_\tau[R^{CLI}_\tau(\bp_0, y; \widetilde{\Delta \bp})^n] &= \int_0^1 R^{CLI}_\tau(\bp_0, y; \widetilde{\Delta \bp})^n d \tau \\
            &= \int_0^1 \left[K_{w, \tau}(\bp_0, y ; \widetilde{\Delta \bp})^n + n K_{w, \tau}(\bp_0, y ; \widetilde{\Delta \bp})^{n-1} S_\tau^{CLI}(\bp_0, y; \widetilde{\Delta \bp})
            \right] d\tau + O(||\widetilde{\Delta \bp}||^{n+2}) \\
            &= \mathfrak{M}_n^{CLI}(\bp_0, y; \widetilde{\Delta \bp}) + \mathfrak{B}_n^{CLI}(\bp_0, y; \widetilde{\Delta \bp}) +O(||\widetilde{\Delta \bp}||^{n+2}),
        \end{split}
    \end{equation*}
    where the last equality follows from Lemma~\ref{lem:hoderleinlemma}. That is, the $n$th moment of $R^{CLI}_\tau(\bp_0, y; \widetilde{\Delta \bp})$ coincides with that of the CLI (as derived in Theorem~\ref{thm:linearallmoments}), up to order $n+1$.

\subsection{Proof of Proposition~\ref{cor:decomposition}}

We first consider a population with homogeneous preferences.
    Since homogeneity implies that ${W}_n(\bp_0, y; \bt) = {W}_1(\bp_0, y; \bt)^n$, in this case the overall behavioral effect amounts to 
    \begin{equation*}
        \begin{split}
            &\frac{1}{2} \left(D_{\widetilde{\bp}} W_1(\bp_0, y; \widetilde{\Delta \bp})^n\widetilde{\Delta \bp} + \frac{n}{n+1} D_{\widetilde{y}} W_{1}(\bp_0, y; \widetilde{\Delta \bp})^{n+1} + nW_{1}(\bp_0, y; \widetilde{\Delta \bp})^{n+1}\right) \\
            &\quad = \mathfrak{D}_{n, 1}(\bp_0, y; \widetilde{\Delta \bp}) + \mathfrak{D}_{n, 2}(\bp_0, y; \widetilde{\Delta \bp}). 
        \end{split}
    \end{equation*}
    This may further decomposed into components due to homothetic and nonhomothetic preferences, respectively. Since homotheticity implies that $D_y w_q^\omega(\bp, y; \bt) = 0$, and as a consequence $D_y W_n(\bp, y; \bt) = 0$, we write that
    \begin{equation*}
        \begin{split}
            \mathfrak{D}_{n, 1}(\bp_0, y; \widetilde{\Delta \bp}) &= \frac{1}{2} \left(D_{\widetilde{\bp}} W_1(\bp_0, y; \widetilde{\Delta \bp})^n\widetilde{\Delta \bp} + nW_{1}(\bp_0, y; \widetilde{\Delta \bp})^{n+1}\right), \\
            \mathfrak{D}_{n, 2}(\bp_0, y; \widetilde{\Delta \bp}) &= \frac{1}{2}\left(\frac{n}{n+1} D_{\widetilde{y}} W_{1}(\bp_0, y; \widetilde{\Delta \bp})^{n+1}\right).
        \end{split}
    \end{equation*}

    We now consider a population with heterogeneous preferences. The additional contribution of heterogeneity amounts to 
    \begin{equation*}
        \begin{split}
            &\mathfrak{B}_n^{CLI}(\bp_0, y; \widetilde{\Delta \bp}) - \left[\mathfrak{D}_{n, 1}(\bp_0, y; \widetilde{\Delta \bp}) + \mathfrak{D}_{n, 2}(\bp_0, y; \widetilde{\Delta \bp})\right]\\
            &\quad = \frac{1}{2}\left(D_{\widetilde{\bp}} \overline{W}_n(\bp_0, y; \widetilde{\Delta \bp})\widetilde{\Delta \bp} + \frac{n}{n+1} D_{\widetilde{y}} \overline{W}_{n+1}(\bp_0, y; \widetilde{\Delta \bp}) + n\overline{W}_{n+1}(\bp_0, y; \widetilde{\Delta \bp})\right) \\
            &\quad =\mathfrak{D}_{n, 3}(\bp_0, y; \widetilde{\Delta \bp}) + \mathfrak{D}_{n, 4}(\bp_0, y; \widetilde{\Delta \bp}).
        \end{split}
    \end{equation*}
    Similar as before, this may further decomposed into components due to homothetic and nonhomothetic preferences, respectively:
    \begin{equation*}
        \begin{split}
            \mathfrak{D}_{n, 3}(\bp_0, y; \widetilde{\Delta \bp}) &= \frac{1}{2}\left(D_{\widetilde{\bp}} \overline{W}_n(\bp_0, y; \widetilde{\Delta \bp})\widetilde{\Delta \bp} + n\overline{W}_{n+1}(\bp_0, y; \widetilde{\Delta \bp})\right), \\
            \mathfrak{D}_{n, 4}(\bp_0, y; \widetilde{\Delta \bp}) &= \frac{1}{2}\left(\frac{n}{n+1} D_{\widetilde{y}} \overline{W}_{n+1}(\bp_0, y; \widetilde{\Delta \bp})\right).
        \end{split}
    \end{equation*}

\clearpage
\section{Additional results}\label{app:additionalresults}

\subsection{Boundedness of consumer welfare}\label{app:boundedness}
We establish that consumer welfare is bounded when income effects are uniformly bounded based on the argument developed by \citet{hausman2016individual} and reviewed in \citet{Nonparametricwelfarereview}. Income effects are said to be uniformly bounded when there exists a finite constant $b$ such that 
\begin{equation*}
    b \geq \max |D_y \bq^\omega(\bp, y)|,
\end{equation*}
where $|\cdot|$ denotes the component-wise absolute value and the maximum is taken across all components. 

We present the argument for the CV; the case for the CLI proceeds analogously. Building on the insights of \citet*{hausman1981exact} and \citet*{vartia1983efficient}, we characterize the CV as the solution to a first-order nonlinear ordinary differential equation. Let $\bp(s):[0,1] \rightarrow \mathcal{P}$ denote a continuous price path with $\bp(0) = \bp_0$ and $\bp(1) = \bp_1$. For simplicity, we adopt the linear path $\bp(s) = \bp_0 + s \Delta \bp$.\footnote{By Slutsky symmetry, the choice of path is immaterial.} Further, define
\begin{equation*}
    w^\omega(s) = e^\omega(\bp(s), v^\omega_0) - y, \quad s \in [0, 1],
\end{equation*}
which measures the CV up to point $s$ along the price path. Differentiating with respect to $s$ yields
\begin{equation*}
    \begin{split}
        D_s w^\omega(s) &= D_\bp e^\omega(\bp(s), v^\omega_0) D_s \bp(s), \quad s \in [0, 1], \\
        &= D_\bp e^\omega(\bp(s), v^\omega_0) \Delta \bp.
    \end{split}
\end{equation*}
By Shephard's lemma, the right-hand side simplifies to $\bq^\omega(\bp(s), y + w^\omega(s))^\intercal \Delta \bp$, so the differential equation becomes
\begin{equation*}\label{eq:nlde}
    \begin{split}
        D_s w^\omega(s) = \bq^\omega(\bp(s), y + w^\omega(s))^\intercal \Delta \bp, \quad s \in [0, 1],
    \end{split}
\end{equation*}
with initial condition $w^\omega(0) = 0$. Hence, the CV is given by $w^{\omega}(1)$, the solution at $s=1$. Existence and uniqueness of the solution follow from standard conditions, such as Lipschitz continuity of individual demand. The corresponding integral form of the differential equation is
\begin{equation*}
    w^\omega(s) = \int_{0}^s \bq^\omega(\bp(r), y + w^\omega(r))^\intercal \Delta \bp dr.
\end{equation*}

Since income effects are uniformly bounded, it follows that for each $s \in [0,1]$
\begin{equation*}
    |\bq^\omega(\bp(s), y + w^\omega(s))^\intercal \Delta \bp| \leq |\bq^\omega(\bp(s), y)^\intercal \Delta \bp| + b |\mathbf{1}^\intercal \Delta \bp| |w^\omega(s)|,
\end{equation*}
where $\mathbf{1}$ is a vector of ones of appropriate dimension. Taking absolute values in the integral form and applying this bound yields:
\begin{equation*}
    \begin{split}
        |w^\omega(s)| &= \Bigg |\int_{0}^s \bq^\omega(\bp(r), y + w^\omega(r))^\intercal \Delta \bp dr \Bigg | \\
        &\leq \int_{0}^s |\bq^\omega(\bp(r), y + w^\omega(r))^\intercal \Delta \bp | dr \\
        & \leq \int_{0}^s  \Big[ |\bq^\omega(\bp(r), y)^\intercal \Delta \bp| + b |\mathbf{1}^\intercal \Delta \bp| |w^\omega(r)| \Big] dr \\
        &= \alpha^\omega(s) + \int_0^s \beta^\omega(r) |w^\omega(r)|dr,
    \end{split}
\end{equation*}
where $\alpha^\omega(s) := \int_{0}^s  |\bq^\omega(\bp(r), y)^\intercal \Delta \bp| dr $ and $\beta^\omega(s) := b |\mathbf{1}^\intercal \Delta \bp|$. Since $\alpha^\omega(s)$ is non-decreasing and  $\beta^\omega(s)$ is non-negative, direct application of Gr\"onwall's inequality gives
\begin{equation*}
    |w^\omega(s)| \leq \alpha^\omega(s) \exp\left(\int_0^s \beta^\omega(r) dr\right).
\end{equation*}
Evaluating at $s=1$, we obtain a finite upper bound on $|w^\omega(1)|$, establishing that the CV is bounded under the assumption of uniformly bounded income effects.

\subsection{Nonidentification of higher-order income effects}
\begin{lem}\label{lem:nonidentification}
    $\mathbb{E}[q^\omega(p,y)( D_y q^\omega(p,y))^n ]$ is not identified from cross-sectional data, for $n \geq 2$.
\end{lem}
\begin{proof}
    We establish nonidentification by providing a counterexample. Suppose individual demand is linear in price and income:
    \begingroup\makeatletter\def\f@size{10}\check@mathfonts
    \begin{equation*}
        q^\omega(p,y) = \omega_1 - p + \omega_2 y,
    \end{equation*}
    \endgroup
    where $\omega_1 \sim U(0,1)$, and $\Pr\left[\omega_2 = 1/3\right] = \Pr\left[\omega_2 = 2/3\right] = 1/2$. \citet*{hausman2016individual} show that for $y < 3$, an observationally equivalent specification is given by the quantile demand function:
    \begingroup\makeatletter\def\f@size{10}\check@mathfonts
    \begin{equation*}
        \wtq^{\wtomega} (p,y) =
        \begin{cases}
            -p + \indic[y < 6\wtomega](y/2 + \wtomega) + \indic[y \geq 6\wtomega] (y/3 + 2\wtomega), & \wtomega \leq 1/2, \\
            -p + \indic[y < 6(1-\wtomega)](y/2 + \wtomega) + \indic[y \geq 6(1- \wtomega)] (2y/3 + 2\wtomega-1), & \wtomega > 1/2, \\
        \end{cases}
    \end{equation*}
    \endgroup
    where $\wtomega \sim U(0,1)$.
    
    For the budget set $(p,y) = (1, 2)$, straightforward calculations yield
    \begingroup\makeatletter\def\f@size{10}\check@mathfonts
    \begin{equation}\label{eq:nonident1}
        \begin{split}
            \mathbb{E}\left[q^\omega(p,y)\left( D_y q^\omega(p,y)\right)^n \right] &= \mathbb{E}[(\omega_1 - p + \omega_2 y)\omega_2^n \mid p = 1, y = 2] \\
            &= (\mathbb{E}\left[\omega_1\right] - 1)\mathbb{E}\left[\omega_2^n\right] + 2 \mathbb{E}\left[\omega_3^n\right] \\
            &= -1/4 [(1/3)^n + (2/3)^n] + [(1/3)^{n+1} + (2/3)^{n+1}] \\
            &= 1/12 (1/3)^n + 5/12 (2/3)^n.
        \end{split}
    \end{equation}
    \endgroup
    However, differentiating the quantile demand function with respect to income yields
    \begingroup\makeatletter\def\f@size{10}\check@mathfonts
    \begin{equation*}
        \wtq^{\wtomega}(p,y)\left(D_y \wtq^{\wtomega} (p,y)\right)^n  \mid_{p = 1, y = 2} =
        \begin{cases}
            \begin{split}
                \indic[1/3 < \wtomega] \wtomega (1/2)^n + \indic[1/3 \geq \wtomega](-1/3 + 2\wtomega) (1/3)^n
            \end{split},
            & \wtomega \leq 1/2, \\
            \begin{split}
                \indic[2/3 > \wtomega]\wtomega (1/2)^n + \indic[2/3 \leq \wtomega](-2/3 + 2\wtomega) (2/3)^n
            \end{split}, 
            & \wtomega > 1/2. \\
        \end{cases}
    \end{equation*}
    \endgroup
    Evaluating the expectation gives
    \begingroup\makeatletter\def\f@size{10}\check@mathfonts
    \begin{equation}\label{eq:nonident2}
        \begin{split}
            \mathbb{E}\left[\wtq^{\wtomega}(p,y)\left(D_y \wtq^{\wtomega} (p,y)\right)^n\right] &= (1/3)^n\int_0^{1/3} (-1/3 + 2\wtomega) + (1/2)^n\int_{1/3}^{1/2} \wtomega \\
            &\quad + (1/2)^n\int_{1/2}^{2/3} \wtomega + (2/3)^n\int_{2/3}^{1} (-2/3 + 2\wtomega) \\
            &= 1/6(1/2)^n + 1/3(2/3)^n.
        \end{split}
    \end{equation}
    \endgroup
    Equations \eqref{eq:nonident1} and \eqref{eq:nonident2} are equal only for $n =1$. Since two observationally equivalent models generate different expectations for $n\geq 2$, $\mathbb{E}[q^\omega(p,y)( D_y q^\omega(p,y))^n ])$ is not identified for $n\geq 2$. 
\end{proof}

\subsection{Comparison to other approaches}\label{sec:comparisons}
We now compare our approximation to two established methods for estimating consumer welfare. For simplicity, we focus on the average CV in a two-good setting. To streamline notation, we denote our second-order approximation with heterogeneous agents (HA) by
\begin{equation*}
    \overline{CV}_{HA}(p_0,p_1, y) = \Delta p M_1(p_0,y) + \frac{(\Delta p)^2}{2}\left(D_p M_1(p_0,y)+\frac{1}{2}D_y M_2(p_0,y)\right).
\end{equation*}
    
\paragraph{Representative agent approach.}
The representative agent (RA) approach assumes that the whole cross section of demand is generated by a single type of consumer. Under this assumption, moments of demand satisfy $M_n(p_0, y) = M_1(p_0, y)^n$. Substituting into the second-order expansion yields the RA approximation:
\begin{equation*}
    \begin{split}
        \overline{CV}_{RA}(p_0, p_1, y) &= \Delta p M_1(p_0,y) + \frac{(\Delta p)^2}{2}\left(D_p M_1(p_0,y)+\frac{1}{2}D_y M_1(p_0,y)^2\right).
    \end{split}
\end{equation*}
The difference between the heterogeneous-agent and representative-agent approximations is
\begin{equation*}
    \begin{split}
        \overline{CV}_{HA}(p_0,p_1, y) - \overline{CV}_{RA}(p_0, p_1, y) &= \frac{(\Delta p)^2}{4} \left(D_y M_2(p_0,y) -  D_y M_1(p_0,y)^2 \right) \\
        &= \frac{(\Delta p)^2}{2} \Cov\left(q^\omega(p_0, y), D_y q^\omega(p_0, y)\right),
    \end{split}
\end{equation*}
which depends on the covariance between consumption and the marginal propensity to consume. Intuitively, in the spirit of \citet{lewbel2001demand}, through conditional heteroscedasticity, our approach corrects for average demand not being compatible with a representative agent.

\paragraph{Bounds-based approach.}
An alternative strategy, when information about the magnitude of income effects is available, is to adopt a bounds-based approach. \citet{hausman2016individual} pursue this route by bounding welfare effects using only the first moment of demand.

To see how our approximation compares, suppose that individual income effects satisfy the uniform bounds $b^l \leq D_y q^\omega(p_0, y) \leq b^u$  for all $\omega$. Define 
\begin{equation*}
    \begin{split}
        CV_B^\omega(p_0,p_1,y; b) &= \Delta p q^\omega(p_0,y)+\frac{(\Delta p)^2}{2}\left[D_p q^\omega(p_0,y) + bq^\omega(p_0,y) \right],
    \end{split}
\end{equation*}
such that after taking expectations
\begin{equation}\label{eq:linbounds}
    \begin{split}
        \overline{CV}_B(p_0,p_1,y; b) &= \Delta p M_1(p_0,y) +\frac{(\Delta p)^2}{2}\left[D_p M_1(p_0,y) + b M_1(p_0,y) \right].
    \end{split}
\end{equation}
Since $D_y q^\omega(p_0, y) \in [b^l, b^u]$ by assumption, it follows directly that $\overline{CV}_B(p_0,p_1,y; b^l) \leq \overline{CV}_{HA}(p_0,p_1,y) \leq \overline{CV}_B(p_0,p_1,y; b^u)$.

How informative are these bounds in practice? If the good is assumed to be normal, the worst-case bounds are such that the percentage point gap between the lower and upper bound amounts to approximately half of the price increase.\footnote{The upper bound to the income effect follows from the budget constraint.} Indeed, from Equation~\eqref{eq:linbounds}, it follows that
    \begin{equation*}
    \frac{{\overline{CV}}_B(p_0,p_1,y; \frac{1}{p}) - {\overline{CV}}_B(p_0,p_1,y; 0)}{{\overline{CV}}_B(p_0,p_1,y; 0)} = \frac{\frac{(\Delta p)^2}{2} \frac{1}{p} M_1(p_0, y)}{\Delta p M_1(p_0, y) + \frac{(\Delta p)^2}{2} D_y M_1(p_0, y)} \approx \frac{1}{2}\frac{\Delta p}{p}.
\end{equation*}
Note that when the difference between the bounds is compared with respect to the first-order effect $\Delta p M_1(p_0, y)$, the result is exact.

\clearpage
\section{Empirical application}\label{app:data}
\subsection{Data construction} 

We use NielsenIQ consumer panel data from the years 2019–2022. Because NielsenIQ revised its product categorization between 2020 and 2021, we adopt the classification scheme used prior to 2021 for consistency. As such, the departments are Health and Beauty Aids (0), Dry Grocery (1), Frozen Foods (2), Dairy (3), Deli (4), Packaged Meat (5), Fresh Produce (6), Non-Food Grocery (7), Alcohol (8), General Merchandise (9), and Magnet Data Products (99). We focus our analysis on food categories and therefore keep departments $1$-$6$ and $8$. We keep purchases and trips information that fall within any of these categories across households and years. For the years $2021$-$2022$, we use module codes associated with Universal Product Codes (UPCs) to map them to department codes used prior to $2021$. For each year of the panel, we compute unit prices paid after discounts for each UPC purchased and for each household. Then, we remove prices paid below the 1st percentile and above the 99th percentile. This removes zero prices that sometimes occur due to discounts and prices that are likely the result of entry mistakes (e.g., price paid of 999.99).

Next, for each household, we aggregate prices paid and quantities purchased to a monthly basis. In particular, prices are aggregated according to the Stone-Lewbel price index proposed by \cite{HoderleinMihaleva2008}. Let $n$ denote the number of departments  and $n_i$ denote the number of UPCs in department $i$. Also let $p_{ij}$ and $q_{ij}$ denote the price and quantity of UPC $j$ in department $i$, respectively. Let $y_i$ denote the expenditure on department $i$ such that $w_{ij} := p_{ij}q_{ij}/y_i$ denote the expenditure share of UPC $j$ in department $i$. Then, the Stone-Lewbel price index is given by
\begin{equation*}
    P_{it} = \frac{1}{k_{it}} \prod_{j=1}^{n_i} \left( \frac{p_{ijt}}{w_{ijt}} \right)^{w_{ijt}},
\end{equation*}

\noindent
where $k_{it}$ is a scaling factor given by
\begin{equation*}
    k_{it} := \prod_{j=1}^{n_i} \overline{w}_{ijt}^{-\overline{w}_{ijt}},
\end{equation*}

\noindent
where $\overline{w}_{ijt}$ represents the average expenditure share of good $j$ in department $i$ and month $t$.

Finally, we construct our dataset by pooling each panel year such that an observation corresponds to a unique household-year-month combination. To account for household economies of
scale, we equivalize household income and household total expenditures according to a modified OECD equivalence scale. Specifically, the equivalized income (EI) is computed as: 
\begin{equation*}
    \text{EI} = I/(1 + 0.5\cdot \mathds{1}(nadults = 2) + 0.3\cdot nchilren),
\end{equation*}

\noindent
where $I$ is household annual income, $\mathds{1}(\cdot)$ is the indicator variable taking value one if the term inside the parentheses is true, $nadults$ is the number of adults, and $nchildren$ is the number of children. The same formula is used for equivalized total expenditures. To mitigate the influence of outliers, we remove observations that are in the bottom and top $2.5\%$ of shares and log prices within each good category, as well as the bottom and top $2.5\%$ of log equivalized expenditures. We further drop any observation with missing information on these variables such that each observation contains purchases from each category of goods. The final estimation sample comprises a total of $269,593$ observations.

\subsection{Inflation}
\begin{figure}[h]
    \centering
        \includegraphics[width=0.75\textwidth]{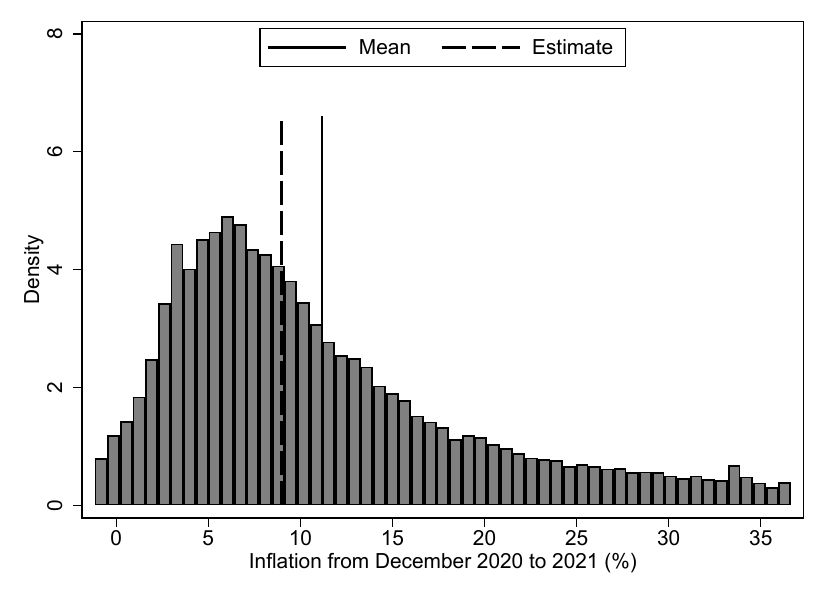}
                \caption{Distribution of inflation as measured by the Fisher index}
        \label{fig:distinflation}
        \captionsetup{justification=justified, singlelinecheck=false}
        \caption*{{\footnotesize \textit{Notes:} The distribution is obtained from household-specific fixed base Fisher indices from UPC-level data between December $2020$ and December $2021$. In particular, we use household-specific expenditure weights and common prices in the computation of the Fisher indices.}\setstretch{1.0} }
\end{figure}

In this section, we take advantage of the panel structure and the UPC-level data available in the NielsenIQ to compare our CLI estimates with inflation rates obtained from a standard Fisher index. As in our application, we focus on the inflation rate from December $2020$ to December $2021$. Then, we compute household-specific inflation rates using the same approach as in \cite{jaravel2020}. Figure~\ref{fig:distinflation} displays the distribution of inflation rates obtained from that panel-based approach.

Figure~\ref{fig:distinflation} reveals a sizable heterogeneity in inflation rates across households. Importantly, the average inflation rate indicated by the vertical solid black line is close to our average CLI indicated by the vertical dashed line, with a mere difference of about $2$ percentage points. It is worth noting that a difference in the cost of living is expected due to sample differences between the sample in our application and this panel-based sample. Indeed, Fisher inflation rates require households to be present in both $2020$ and $2021$, whereas our application works with only cross-sectional data. Furthermore, the contribution of a UPC to the Fisher inflation rate only arises from observations where a UPC is purchased in both periods. This condition is not always met because households may not purchase the same UPC in December $2020$ and December $2021$. In spite of sample differences, both approaches yield similar inflation rates for the period under consideration.

Next, we investigate how inflation rates vary with demographics by regressing Fisher inflation rates on demographics. Table \ref{table:OLS} reports coefficients from separate regressions of household-specific inflation rates on categorical indicators for household size, household income, and race. In each instance, the reference group is the first variable of the demographic group. 

\begin{table}[htbp]\centering
\caption{Regressions of inflation on demographic variables}
\label{table:OLS}
\def\sym#1{\ifmmode^{#1}\else\(^{#1}\)\fi}

\begin{adjustbox}{max width=\textwidth, center}

\begin{threeparttable}[b]

\begin{tabular}{l*{8}{c}}
\toprule
                &\multicolumn{1}{c}{ Dry Grocery}&\multicolumn{1}{c}{Frozen Foods}&\multicolumn{1}{c}{Dairy}&\multicolumn{1}{c}{Deli}&\multicolumn{1}{c}{ Packaged Meat}&\multicolumn{1}{c}{Fresh Produce}&\multicolumn{1}{c}{Alcohol}&\multicolumn{1}{c}{All}\\
\midrule
Constant          &    6.459\sym{***}&    15.15\sym{***}&    3.192\sym{***}&    26.92\sym{***}&    9.915\sym{***}&    4.952\sym{***}&    2.702\sym{***}& 11.075\sym{***} \\
                & (0.0600)         &  (0.189)         & (0.0975)         &  (0.295)         &  (0.211)         & (0.0552)         &  (0.229)            & (0.103) \\
    \textit{Household size} &                 &                 &                 &                 &                 &                 &                 \\
\quad 2               &  -0.0577         &    1.639\sym{***}&    0.109         &   -0.874\sym{***}&   0.0784         &  0.00632         &    0.294   & -0.052      \\
                & (0.0619)         &  (0.194)         & (0.0994)         &  (0.302)         &  (0.214)         & (0.0562)         &  (0.229)    & (0.107)     \\
\quad 3+               &   0.0207         &    0.781\sym{***}&   0.0227         &   -1.272\sym{***}&    0.435\sym{*}  &  -0.0163         &    0.661\sym{**}   &  -0.256\sym{**} \\
                & (0.0671)         &  (0.209)         &  (0.108)         &  (0.331)         &  (0.225)         & (0.0621)         &  (0.261)   &  (0.117)      \\
    \textit{Household income} &                 &                 &                 &                 &                 &                 &                 \\
\quad $25$k–$35$k             &  -0.0521         &   -0.138         & -0.00893         &    0.296         &    0.260         &  0.00129         &   -0.150         & -0.061 \\
                & (0.0676)         &  (0.205)         &  (0.104)         &  (0.341)         &  (0.206)         & (0.0620)         &  (0.252)            & (0.120) \\
\quad $35$k-$50$k             & -0.00510         &   -0.105         &  -0.0742         &    0.176         &   -0.102         &  -0.0606         &   -0.216         & 0.0142 \\
                & (0.0599)         &  (0.181)         & (0.0916)         &  (0.298)         &  (0.181)         & (0.0547)         &  (0.221)           & (0.106) \\
\quad $50$k-$70$k            &  -0.0793         &    0.158         &   -0.174         &   -0.313         &   -0.208         &   -0.132\sym{**} &   0.0943         & -0.228\sym{*} \\
                & (0.0719)         &  (0.218)         &  (0.109)         &  (0.353)         &  (0.219)         & (0.0636)         &  (0.245)            & (0.127) \\
\quad $>70$k       &   -0.340\sym{**} &   0.0300         &   -0.152         &   -1.568\sym{**} &   -1.360\sym{***}&  -0.0411         &    0.770         & -0.645\sym{***} \\
                &  (0.143)         &  (0.466)         &  (0.230)         &  (0.659)         &  (0.517)         &  (0.127)         &  (0.518)            &  (0.243) \\
    \textit{Race}   &                 &                 &                 &                 &                 &                 &                 \\
\quad Black           &   -0.260\sym{***}&    0.988\sym{***}&   -0.152         &    1.832\sym{***}&    0.725\sym{***}&    0.247\sym{***}&    0.257         &  1.499\sym{***} \\
                & (0.0749)         &  (0.224)         &  (0.125)         &  (0.351)         &  (0.226)         & (0.0693)         &  (0.279)            & (0.131) \\
\quad Asian           &   -0.496\sym{***}&    3.931\sym{***}&   -0.637\sym{***}&    2.351\sym{***}&   -0.536         &    0.217\sym{**} &    1.576\sym{***}&   1.886\sym{***} \\
                &  (0.129)         &  (0.374)         &  (0.214)         &  (0.528)         &  (0.455)         &  (0.101)         &  (0.586)            & (0.212)\\
\quad Other           &   -0.293\sym{***}&    0.781\sym{**} &   -0.115         &    2.610\sym{***}&    0.624\sym{*}  &  0.00632         &    0.107         &  0.451\sym{**} \\
                &  (0.110)         &  (0.340)         &  (0.174)         &  (0.548)         &  (0.350)         & (0.0993)         &  (0.400)            &  (0.192)  \\
\midrule
Observations    &    34201         &    19518         &    16791         &    13963         &     9236         &    25924         &     3958         & 41318 \\
\bottomrule
\end{tabular}

  \begin{tablenotes}
    \item[*] {\footnotesize To account for differences in household composition, annual income is equivalized using a modified OECD scale. The equivalized income (EI) is computed as: EI $= I/(1 + 0.5\cdot \mathds{1}(nadults = 2) + 0.3\cdot nchilren)$, where $I$ is household annual income, $\mathds{1}(\cdot)$ is the indicator variable, $nadults$ is the number of adults, and $nchildren$ is the number of children.}
    \item[*] {\footnotesize Standard errors in parentheses. Significance levels: $p < 0.10$ (\sym{*}), $p < 0.05$ (\sym{**}), $p < 0.01$ (\sym{***}).}  \setstretch{1.0}
    
  \end{tablenotes}

\end{threeparttable}

\end{adjustbox}

\end{table}


Table \ref{table:OLS} shows that, relative to single-adult households, those with two adults show significantly higher inflation for frozen foods and significantly lower inflation for deli products. For households with three or more adults, frozen food inflation is again significantly higher, while deli inflation is lower. Notably, these differences are not uniformly significant across all food categories. Next, higher-income households generally face lower inflation, particularly in more processed or higher-end categories. For instance, households earning above $70$k dollars experience significantly lower inflation in dry groceries, deli, and packaged meat. A similar (though weaker) pattern is observed for the $50$k–$70$k group. This suggests a potential substitution toward less inflationary goods or price resilience among wealthier households. Finally, relative to White households, Black households experience significantly higher inflation in frozen foods, deli, packaged meat, and fresh produce. Asian households also face notably higher inflation in frozen foods, deli, and alcohol, but significantly lower inflation in dry groceries and dairy. These results point to substantial heterogeneity in inflation experiences across racial groups, likely reflecting differences in consumption patterns or access to lower-priced alternatives.

The final column ("All") aggregates inflation across all categories, showing that larger households and higher-income households tend to experience less inflation on average, while racial minority groups, such as Black and Asian households, face significantly higher inflation. These results emphasize the unequal burden of inflation across demographics.

\end{document}